\documentstyle[12pt]{report}

 \catcode`\@=11 

\newcommand{\nonumsection}[1] {\vspace{12pt}\noindent{\bf #1}
        \par\vspace{5pt}}

\def\thebibliography#1{\nonumsection{\large \it References}\list
  {[\arabic{enumi}]}{\settowidth\labelwidth{[#1]}\leftmargin\labelwidth
    \advance\leftmargin\labelsep
    \usecounter{enumi}}
    \def\newblock{\hskip .11em plus .33em minus .07em}
    \sloppy\clubpenalty4000\widowpenalty4000}

\newcommand{\tcaption}[1]{
        \addtocounter{table}{1}
{{\tenrm\offinterlineskip Table~\thetable . #1} }\hfil\break }

\newcommand{\fcaption}[1]{
        \addtocounter{figure}{1}
         {{\tenrm Fig.~\thefigure . #1} }\hfil\break }

\newcommand{\be}{\begin{eqnarray}}
\newcommand{\ee}{\end{eqnarray}}
\newcommand{\dslash}{\partial \hskip -0.5em /}
\newcommand{\Dslash}{D \hskip -0.7em /}
\newcommand{\tr}{{\rm tr}}
\newcommand{\Tr}{{\rm Tr}}

\newcommand{\La}{{\cal L}}
\newcommand{\A}{{\cal A}}

\newcommand{\T}{{\cal T}}
\newcommand{\ie}{{\it i.e.}\ }
\newcommand{\eg}{{\it e.g.}\ }

\input psfig

\begin{document}

\rightline{UNITU-THEP-14/1993}
\rightline{October 1993}
\rightline{hep-ph/9310309}

\vskip 1.5truecm
\centerline{\Large\bf Hyperons in the Bound State Approach to the }
\vskip .3cm
\centerline{\Large\bf Nambu--Jona-Lasinio Chiral Soliton $^\dagger $}
\vskip 1.0cm
\centerline{ H.\ Weigel$^\ddagger$, R.\ Alkofer and H.\ Reinhardt }
\vskip .3cm
\centerline{Institute for Theoretical Physics}
\centerline{T\"ubingen University}
\centerline{Auf der Morgenstelle 14}
\centerline{D-72076 T\"ubingen, Germany}
\vskip 1.5cm
\centerline{\bf ABSTRACT}

\vskip 0.5cm
\noindent
For the Nambu--Jona--Lasinio (NJL) model in the proper time
regularization scheme the chiral soliton is investigated for
the case of three flavors within the framework of the bound state
approach to strangeness. For this purpose meson fluctuations off
the non--strange soliton are considered. For the kaon P--wave
channel the emerging Bethe--Salpeter equation is solved yielding
the bound state energy and wave--function. This kaon bound state
also induces a strange valence quark wave--function. Collective
coordinates are introduced for the SU(2) isorotations and the coupling
of the bound state to these rotations is determined. The total spin
is decomposed into parts carried by the soliton and the bound state.
The absolute value of the former is identical to the total isospin
while the latter is demonstrated to be identical to the expectation
value of the grand spin, the sum of spin and isospin. The functional
trace involves quark states with various grand spins which get
polarized by the kaon bound state. This polarization results in a
normalization of the coupling between the collective rotation and
the bound state different from the Skyrme model. Nevertheless after
quantization the expression for the baryon masses can be cast into
the form found by Callan and Klebanov for the Skyrme model. Numerical
results for the baryon spectrum are compared to those obtained in the
same model using the collective approach of Yabu and Ando. It is found
that for $\frac{1}{2}^+$ baryons the bound state treatment reproduces
the experimental data better than the collective approach while there
are only small differences between the two approaches for the
$\frac{3}{2}^+$ baryons. Both treatments underestimate the mass
splittings for baryons with different strangeness. This short-coming
is conjectured to be inherited from the meson sector of the NJL model
where too small a ratio of the kaon and pion decay constants is predicted.
\vfill

\noindent
------------\hfil\break
\noindent
$^\dagger $ Supported by the Deutsche Forschungsgemeinschaft (DFG) under
contract Re 856/2-1.
\newline
\noindent
$^\ddagger$ Supported by a Habilitanden--scholarship of the DFG.
\eject

\stepcounter{chapter}
\leftline{\large\it 1. Introduction}

For a large number of colors QCD is known to reduce to an effective theory
of weakly interacting mesons \cite{tho74}. Witten conjectured
\cite{wi79} that within this effective theory baryons emerge as soliton
solutions. Although Witten's conjecture has never been proved, the
soliton picture of baryons has turned out quite successful in recent years.

Since the effective meson theory emerging from QCD is not explicitly
known, one has resorted to phenomenological effective meson theories,
which possess soliton solutions. The most popular ones perhaps are
the Skyrme model and the gauged $\sigma$-model. Investigations within
these models have satisfactorily explained the wealth of spectroscopic
baryon data.  See ref. \cite{ho93} for a recent compilation of
references on soliton models for baryons.

Many attempts have been made to derive the effective meson theory
from QCD. These attempts have shown that at low energies the effective
meson theory is almost entirely determined by chiral symmetry. This
suggests to study the bosonization in simpler chirally invariant models
of the quark flavor dynamics. The prototype of these models is the
Nambu--Jona--Lasinio (NJL) model \cite{na61}. This model can exactly be
bosonized and the resulting effective meson theory is in satisfactory
agreement with the low-energy data for pseudoscalar and (axial-) vector
mesons \cite{eb86}. Recently, also the self-consistent soliton solutions
to the bosonized NJL model have been found for the case of two flavors
\cite{re88,me89,al90}. These soliton calculations give a transparent
explanation of how baryons emerge as solitons in effective meson theories
starting from an underlying quark theory and strongly support
Witten's conjecture \cite{al92,zu93}.

Unfortunately the soliton description of baryons cannot be
straightforwardly extended to also include strangeness since flavor
symmetry breaking plays a significant role. Within the soliton picture
hyperons have been treated in two conceptionally different ways, either
in the SU(3)--flavor collective approach of Yabu and Ando \cite{ya88}
or in the bound state approach of Callan and Klebanov \cite{ca85,ca88}.
In the Yabu-Ando approach SU(3)--flavor is considered to still be an
approximate symmetry. Collective coordinates are then introduced
describing the rotation of the classical soliton in SU(3)--flavor
space. Canonical quantization of these coordinates yields a collective
Hamiltonian which also includes SU(3) symmetry breaking terms. The
special feature of the Yabu--Ando approach is that this Hamiltonian can
be diagonalized exactly. The corresponding eigenstates are identified
as hyperons. Due to the explicit flavor symmetry breaking the associated
eigenvalues are correlated to the hypercharge quantum number and thus
remove the degeneracy of baryons with different strangeness. The
Yabu-Ando approach has already been applied to describe the hyperons
as chiral solitons of the NJL model. Comprehensive studies
are reported in refs. \cite{we92,bl93}. On the other hand
in the Callan-Klebanov approach the flavor symmetry breaking by
the strange quark is considered large implying an asymmetric
treatment of the strange and non-strange sectors. The
hyperons are here described as bound states of a strange meson
({\it i.e.} a kaon in the case of pseudoscalars) in the
background field of the non--strange soliton. Each occupation
of this bound state lowers strangeness by one unit. Collective
coordinates are only introduced for the SU(2) (iso-) rotations,
{\it i.e.} the real zero modes. Treating these coordinates in the
well--known collective manner lifts the degeneracy of baryons with
the same strangeness charge but different spin or isospin as {\it e.g.}
the $\Lambda$ and $\Sigma$ hyperons.

For hyperons both approaches have been applied to various meson
soliton models with similar success. The original Yabu--Ando approach
to the rigid rotator of the Skyrme model has for example been extended
to the slow rotator \cite{schw92}, also the influence of vector mesons
on baryon properties in this treatment has been studied \cite{pa92}.
For similar investigations in the bound state approach the reader may
{\it e.g.} consult refs. \cite{ca88},\cite{sc88}--\cite{pa93}. For
yet heavier baryons containing {\it e.g.} charmed or bottom quarks the
bound state approach should become the more appropriate one and provide
a more realistic picture \cite{je92}--\cite{ri93}.

In the present model we study the hyperons within the NJL model
applying the bound state approach to the bosonized NJL model. Some
preliminary results have already been reported in ref. \cite{we93}.
The organization of the present paper is as follows: In section 2) we
review the treatment of meson fluctuations off the chiral soliton
in the NJL model. In section 3) we specialize to the kaon bound
state in the background field of the soliton. The main effort is
devoted to the semiclassical quantization of the soliton--bound state
system. This is described in section 5). The numerical results
for the baryon mass spectrum are presented in section 6) which
also contains a comparison with the collective approach to hyperons
in the NJL model. Conclusive remarks as well as an outlokk are
given in section 7). We leave several technicalities to three appendices.

\vskip 1.5cm

\stepcounter{chapter}
\leftline{\large\it 2. Fluctuations off the Soliton}

\medskip

The basis of our considerations is the chirally symmetric NJL Lagrangian
\be
\La = \bar q (i\dslash - \hat m^0 ) q +
      2G_{\rm NJL} \sum _{i=0}^{N_f^2-1}
\left( (\bar q \frac {\lambda ^i}{2} q )^2
      +(\bar q \frac {\lambda ^i}{2} i\gamma _5 q )^2 \right) .
\label{NJL}
\ee
Here $q$ denotes the quark field and $\hat m^0$ the current quark mass
matrix. The matrices $\lambda ^i/2$ are the generators of the flavor
group $U(N_f)$. We will consider the case of three flavors ($N_f=3$)
and neglect isospin breaking, \ie the current quark mass matrix acquires
the form $\hat m^0={\rm diag}(m^0,m^0,m^0_s)$ as $m^0=m_u^0=m_d^0$  is
assumed. The NJL coupling constant $G_{\rm NJL}$ will later be
determined from meson properties.

Applying functional integral bosonization techniques the model
(\ref{NJL}) can be rewritten in terms of composite meson fields \cite{eb86}.
The corresponding effective action is given by the sum
$\A_{NJL}=\A_F+\A_m$ of a fermion determinant
\be
\A_F=\Tr\log(i\Dslash)=\Tr\log\left(i\dslash-(P_RM+P_LM^{\dag})\right)
\label{fdet}
\ee
and a purely mesonic part
\be
\A_m=\int d^4x\left(-\frac{1}{4G_{\rm NJL}}
\tr(M^{\dag}M-\hat m^0(M+M^{\dag})+(\hat m^0)^2)\right).
\label{ames}
\ee
Here $P_{R,L}=(1\pm \gamma _5)/2$ denote the projectors onto right--
and left--handed quark fields, respectively. The complex matrix
$M=S+iP$ parametrizes the scalar and pseudoscalar meson fields.

Obviously, the NJL model is non--renormalizable and has to be
regularized. As in original study \cite{re88} of the NJL soliton we
will apply the $O(4)$ invariant proper time regularization \cite{sch51}
to the fermion determinant in Euclidean space. In Euclidean space
the fermion determinant in general is a complex quantity
$\A_F=\A_R+\A_I$. Proper time regularization amounts to replacing
the real part of the fermion determinant $\A_R$ by a parameter integral
\be
\A_R=\frac{1}{2}\Tr\log\left(\Dslash_E^{\dag}\Dslash_E\right)
\longrightarrow -\frac{1}{2}\int_{1/\Lambda ^2}^\infty
\frac{ds}s\Tr\exp\left(-s\Dslash_E^{\dag}\Dslash_E\right) .
\label{arreg}
\ee
Note that (\ref{arreg}) becomes exact in the limit $\Lambda\to
\infty$ up to an irrelevant constant. Although the imaginary part
\be
\A_I=\frac{1}{2}\Tr\log\left((\Dslash_E^{\dag})^{-1}\Dslash_E\right)
\label{af3}
\ee
is finite we will later also regularize $\A_I$ in a way consistent
with (\ref{arreg}) \cite{al92a}.

As the aim of the paper is to investigate hyperons as systems of
bound kaons in the soliton background we have to parametrize the
pseudoscalar meson field in a way which allows for both, the static
soliton and meson fluctuations. In this context it is convenient to
use the following parametrization for the scalar and pseudoscalar
fields \cite{we93}
\be
M=\xi_0\xi_f\Sigma\xi_f\xi_0.
\label{defm}
\ee
The matrix $\Sigma$ is hereby Hermitian whereas the matrices $\xi_0$ and
$\xi_f$ are unitary.  The quantity $\xi_0$ carries the information about
the static soliton while $\xi_f$ denotes the fluctuations.
In the baryon number zero sector $\xi_0$ is replaced by the unit matrix.
Note that this parametrization differs from the usually adopted
unitary gauge $M=\xi_L^\dagger\Sigma\xi_R,\ \xi_L^\dagger=\xi_R$.

Varying the action with respect to $\Sigma$ yields the Schwinger--Dyson
or gap equation which determines the vacuum expectation value (VEV),
$\langle \Sigma \rangle ={\rm diag}(m,m,m_s)$ and relates it to the
quark condensate $\langle \bar q q \rangle _i$
\be
m_i & = & m_i^0+m_i^3\frac{N_C G_{\rm NJL}}{2\pi^2}
\Gamma\left(-1,(\frac{m_i}{\Lambda})^2\right)
=m_i^0-2G_{\rm NJL} \langle \bar q q \rangle _i .
\label{conmass}
\ee
Here $m_i=m,m_s$ denote the constituent masses of non--strange
and strange quarks, respectively.

Space-time dependent fluctuating pseudoscalar meson
fields $\eta_a(x)$ are introduced via
\be
\xi_f(x)={\rm exp}\left(i\sum_{a=1}^8\eta_a(x)\lambda_a/2\right) .
\label{defeta}
\ee
Expanding the effective action up to second order in the fluctuations
$\eta_a(x)$ allows to extract the inverse propagator for the
pseudoscalar mesons $P_{ij}=\sum_{a=1}^{N_f^2-1}\eta^a\lambda^a_{ij}$
\cite{we92}
\be
D_{ij,kl}^{-1}(q^2)=
\Big(-\frac{(m_i^0+m_j^0)(m_i+m_j)}{2G_{\rm NJL}}-\Pi_{ij}(q^2)\Big)
\delta_{il}\delta_{kj}.
\label{prop}
\ee
The polarization operator $\Pi_{ij}(q^2)$ is given by
\be
\Pi_{ij}(q^2)=-2q^2f_{ij}^2(q^2)+2(m_i-m_j)^2f_{ij}^2(q^2)
-\frac{1}{2}(m_i^2-m_j^2)\Big(\frac{\langle\bar q q \rangle_i}{m_i}-
\frac{\langle\bar q q \rangle_j}{m_j}\Big)
\label{polar}
\ee
wherein
\be
f^2_{ij}(q^2)=\frac{1}{4}(m_i+m_j)^2\frac{N_c}{4\pi^2}\int_0^1dx\
\Gamma\Big(0,[(1-x)m_i^2+xm_j^2-x(1-x)q^2]/\Lambda^2\Big).
\label{fq}
\ee
The Bethe--Salpeter equation $D^{-1}(q^2)P=0$ which determines the
physical meson masses $m_{\rm phys}$ is equivalent to the condition
that the meson propagator acquires a pole:
\be
{\rm det}\left(D_{ij,kl}^{-1} (q^2=m_{\rm phys}^2)\right)=0.
\label{BSeq}
\ee
Note that $f^2_{ij}(q^2=m_{\rm phys}^2)$ is the corresponding
on--shell meson decay constant. From eqn. (\ref{fq}) we can read
off the pion decay constant
\be
f_\pi ^2 =m^2 \frac {N_c}{4\pi ^2} \int _0^1 dx
\Gamma\Big(0,[m^2-x(1-x)m_\pi^2]/\Lambda ^2\Big)
\label{fpi}
\ee
as well as the kaon decay constant
\be
f_K^2= \frac 14 (m+m_s)^2 \frac {N_c}{4\pi ^2} \int _0^1 dx
\Gamma\Big(0,[xm^2+(1-x)m_s^2-x(1-x)m_K^2]/\Lambda ^2\Big) .
\label{fK}
\ee
Using eqs.\ (\ref{BSeq},\ref{fpi}) the four parameters of the model
(\ref{NJL}), the coupling constant $G_{\rm NJL}$, the cutoff
$\Lambda $ and the
two current masses $m^0$ and $m_s^0$ may be determined. The current
quark masses are determined by the pion and kaon masses,
$m_\pi=135$ MeV and $m_K=495$ MeV. Fixing now the pion decay constant
$f_\pi = 93$ MeV yields too small a value for the kaon decay constant,
see table 2.1. On the other hand, requiring $f_K=113$ MeV leaves us with
too large a value for $f_\pi$.\footnote{Determining $f_\pi$ or $f_K$
fixes the ratio $\Lambda /m$. This leaves one adjustable parameter,
\eg the coupling constant $G_{\rm NJL}$. However, as $G_{\rm NJL}$ is
not very transparent we will use the gap equations (\ref{conmass}) to
reexpress it in terms of the up constituent mass $m$.}

\tcaption{The up and strange constituent and current masses, the cutoff
and the pion and the kaon decay constant for the parameters used later.}
{}~
\newline
\centerline{
\begin{tabular}{|c|c|c|c|c|c|}
   \hline
$m$ (MeV) & $m_s$ (MeV) & $m^0_s/m^0_u$
& $\Lambda$ (MeV) & $f_\pi$ (MeV) & $f_K$ (MeV) \\
\hline
\hline
350&577&23.5&641&93.0 &104.4\\
\hline
400&613&22.8&631&93.0 &100.3\\
\hline
450&650&22.4&633&93.0 & 97.4\\
\hline
500&687&22.3&642&93.0 & 95.5\\
\hline
\hline
350&575&24.3&698& 99.3&113.0\\
\hline
400&610&23.9&707&103.0&113.0\\
\hline
450&647&23.6&719&105.7&113.0\\
\hline
500&685&23.4&734&107.9&113.0\\
\hline
\end{tabular}}
{}~ \newline

For the investigation of the baryon sector we constrain the meson
fields to the chiral circle, {\it i.e.} we replace $\Sigma$ by
its VEV $\langle \Sigma \rangle$ in eqn. (\ref{defm}).
The Euclidean Dirac operator $\Dslash_E$ may then be written as
\be
i\beta\Dslash_E=-\partial_\tau-h=
-\partial_\tau-\mbox {\boldmath $\alpha \cdot p $}
-\T \beta\left(\xi_f\langle\Sigma\rangle\xi_f P_R
+\xi_f^\dagger\langle\Sigma\rangle\xi_f^\dagger P_L\right)\T^\dagger
\label{de1}
\ee
wherein $\tau=ix_0$ is the Euclidean time and the unitary matrix
\be
\T=\xi_0 P_L + \xi_0^\dagger P_R
\label{deft}
\ee
parametrizes the static soliton. Expanding around the static meson field
configuration the Hamiltonian $h$ in (\ref{de1}) may be written
\be
h=h_{(0)}+h_{(1)}+h_{(2)}+\cdot\cdot\cdot.
\label{de2}
\ee
where the subscripts label the powers of the meson fluctuations $\eta_a$.
We furthermore assume the hedgehog {\it ansatz} for the pseudoscalar
fields
\be
\xi_0(\mbox {\boldmath $x $}) & = & {\rm exp}\left(\frac{i}{2}
{\mbox{\boldmath $\tau$}} \cdot{\bf \hat r}\ \Theta(r)\right).
\label{chsol}
\ee
Defining the grand spin ${\bf G}$ as the vectorial sum of spin
and isospin (see also appendix C) the hedgehog has the celebrated
feature of being invariant under grand spin transformations. With
the hedgehog {\it ansatz} one obtains for $h_{(i)}$\cite{we93}
\be
\!\!h_{(0)}({\bf r})&=&\!\!\mbox {\boldmath $\alpha \cdot p $} +
{\cal T}\beta\langle\Sigma\rangle{\cal T}^\dagger =
{\mbox {\boldmath $\alpha \cdot p $}} +
\pmatrix{m\ {\rm exp}\left(i\gamma_5{\mbox{\boldmath $\tau$}}
\cdot{\bf \hat r}\ \Theta(r)\right) & 0 \cr 0 & m_s\cr}
\label{h0}
\\*
\!\!h_{(1)}({\bf r},t)&=&\!\!i\T\beta\gamma_5
\left(m \mbox {\boldmath $\eta \cdot \tau $}+\frac{1}{2}
(m+m_s)\sum_{\alpha=4}^7\eta_\alpha\lambda_\alpha\right)\T^\dagger
\label{h1}
\\*
\!\!h_{(2)}({\bf r},t)&=&\!\!-\T\beta\left(
\frac{m}{2\sqrt3}(\frac{2}{\sqrt3}+\lambda_8)
\mbox {\boldmath $\eta \cdot \eta $}+\frac{m+m_s}{8}
\Big\{\sum_{\alpha=4}^7\eta_\alpha\lambda_\alpha,
\sum_{\beta=4}^7\eta_\beta\lambda_\beta\Big\}\right)\T^\dagger .
\label{h2}
\ee
The meson field $\eta_8$ does not couple to the soliton and is hence
discarded. Using that the $h_{(i)}$ are hermitian operators and that
$h_{(0)}$ is time independent the expansion of the action up to second
order in the fluctuations is given by \cite{we93}
\be
\A_R &=& \A_R^{(0)}+\A_R^{(1)}+\A_R^{(2)}+\ldots
\label{arexpa} \\*
\A_R^{(0)} &=& -\frac{1}{2}\Tr\int_{1/\Lambda^2}^\infty
\frac{ds}{s}\hat K_0(s)
\label{ar0} \\*
\A_R^{(1)} &=& \frac{1}{2}\Tr\int_{1/\Lambda^2}^\infty ds
\hat K_0(s)\big\{h_{(1)},h_{(0)}\big\}
\label{ar1} \\*
\A_R^{(2)} &=& \frac{1}{2}\Tr\int_{1/\Lambda^2}^\infty ds
\hat K_0(s)\left(\big\{h_{(2)},h_{(0)}\big\}+h_{(1)}^2\right)
-\frac{1}{4}\Tr\int_{1/\Lambda^2}^\infty ds\int_0^s ds^\prime
\hat K_0(s-s^\prime)
\nonumber \\*
&&\qquad \times\left(\big[\partial_\tau,h_{(1)}\big]
\hat K_0(s^\prime)\big[\partial_\tau,h_{(1)}\big]+
\big\{h_{(1)},h_{(0)}\big\}\hat K_0(s^\prime)
\big\{h_{(1)},h_{(0)}\big\}\right).
\label{ar2}
\ee
for the real part and
\be
\A_I=\A_I^{(2)}+\ldots \rightarrow
\Tr\int_{1/\Lambda^2}^\infty ds\int_0^s ds^\prime
\hat K_0(s-s^\prime)\partial_\tau h_{(1)}
\hat K_0(s^\prime)h_{(0)}h_{(1)}+\ldots
\label{aiexpa}
\ee
for the imaginary part.
$\hat K_0(s)={\rm exp}\left(s(\partial_\tau^2-h_{(0)}^2)\right)$
denotes the zeroth--order heat kernel operator \cite{re89}. Note that
we have also regularized the imaginary part \cite{al92a}.

The temporal part of the functional trace can be performed with the help
of a Fourier transformation for Euclidean time
\be
\eta_a({\bf r},-i\tau) & = & \int_{-\infty}^{+\infty}
\frac{d\omega}{2\pi}\tilde \eta_a({\bf r},i\omega)
{\rm e}^{-i\omega\tau}
\label{feta}
\\*
h_{(1)}({\bf r},-i\tau) & = & \int_{-\infty}^{+\infty}
\frac{d\omega}{2\pi}\tilde h_{(1)}({\bf r},i\omega)
{\rm e}^{-i\omega\tau}\quad {\rm and} \nonumber \\*
h_{(2)}({\bf r},-i\tau) & = &
\int_{-\infty}^{+\infty}\frac{d\omega}{2\pi}
\int_{-\infty}^{+\infty}\frac{d\omega^\prime}{2\pi}
\tilde h_{(2)}({\bf r},i\omega,i\omega^\prime)
{\rm e}^{-i(\omega+\omega^\prime)\tau}
\label{fham}
\ee
wherein $\tilde h_{(i)}$ are obtained from $h_{(i)}$ (\ref{h1},\ref{h2})
through substitution of the fluctuating meson fields by their Fourier
transforms (\ref{feta}). The frequency $\omega$ has to be continued
back to Minkowski space in order to obtain physically relevant
results.\footnote{Strictly speaking, the Gaussian integrals which have
to be calculated in order to do the temporal trace are converging in
Minkowski space only with appropriate boundary conditions. Therefore
it is convenient to perform all analytical calculations in Euclidean space.}
The spatial
part of the trace as well as the traces over Dirac and flavor indices
are evaluated using the eigenstates of the static one-particle
Hamiltonian $h_{(0)}$
\be
h_{(0)} | \mu\rangle = \epsilon_\mu | \mu\rangle .
\label{diagham}
\ee
Hereby the eigenstates with strangeness $\pm1$ are free Dirac spinors
whereas those corresponding to the light quarks are functionals of the
soliton field, see Appendix C.

As expected, the zeroth order term in the action (\ref{ar0}) provides
the static soliton energy and the linear term (\ref{ar1}) the
corresponding equation of motion \cite{we93}. This equation, together
with (\ref{diagham}) serves to obtain the classical soliton profile
$\Theta(r)$ self--consistently. In Minkowski space the quadratic term
$\A_F^{(2)}=\A_R^{(2)}+\A_I^{(2)}$ is given by
\be
\A_F^{(2)} & = & \frac{N_C}{2}\int_{1/\Lambda^2}^\infty
\frac{ds}{\sqrt{4\pi s}}\sum_\mu 2\epsilon_\mu
{\rm e}^{-s\epsilon_\mu^2}\int^{+\infty}_{-\infty}
\frac{d\omega}{2\pi}\langle\mu|\tilde h_{(2)}({\bf r},\omega,-\omega)
|\mu\rangle
\nonumber \\*
&&\!\!\!+\frac{N_C}{4}\int_{1/\Lambda^2}^\infty ds \sqrt{\frac{s}{4\pi}}
\sum_{\mu\nu}\int^{+\infty}_{-\infty}\frac{d\omega}{2\pi}
\langle\mu|\tilde h_{(1)}({\bf r},\omega)|\nu\rangle
\langle\nu|\tilde h_{(1)}({\bf r},-\omega)|\mu\rangle
\label{afquad}
\\*
&&\times \left\{\frac{{\rm e}^{-s\epsilon_\mu^2}
+{\rm e}^{-s\epsilon_\nu^2}}{s}
+[\omega^2-(\epsilon_\mu+\epsilon_\nu)^2]
R_0(s;\omega,\epsilon_\mu,\epsilon_\nu)
-4\omega\epsilon_\nu R_1(s;\omega,\epsilon_\mu,\epsilon_\nu)
\right\}.
\nonumber
\ee
The term containing only odd powers of $\omega$ originates from
the imaginary part. The information on the orderings of the
operators in eqns. (\ref{ar2},\ref{aiexpa}) is contained in the
Feynman parameter integrals
\be
R_i(s;\omega,\epsilon_\mu,\epsilon_\nu)
=\int_0^1 x^i dx\ {\rm exp}\left(-s[(1-x)\epsilon_\mu^2
+x\epsilon_\nu^2-x(1-x)\omega^2]\right)
\label{regfct}
\ee
which represent moments of the quark loop in the presence of
the soliton.

Besides the Dirac sea also the explicit occupation of the valence
quark level contributes to the action as long as the associated
energy eigenvalue $\epsilon_{\rm val}$ is positive. The corresponding
contribution to the second order action reads
\be
\A_{\rm val}^{(2)} & = & -\eta^{\rm val}N_C
\int^{+\infty}_{-\infty}\frac{d\omega}{2\pi}
\Big(\langle{\rm val}|\tilde h_{(2)}({\bf r},\omega,-\omega)
|{\rm val}\rangle \nonumber \\*
&&\qquad\qquad\qquad
+\sum_{\mu\ne{\rm val}}\frac{
\langle{\rm val}|\tilde h_{(1)}({\bf r},\omega)|\mu\rangle
\langle\mu|\tilde h_{(1)}({\bf r},-\omega)|{\rm val}\rangle}
{\epsilon_{\rm val}-\omega-\epsilon_\mu}\Big) .
\label{valquad}
\ee
Here $\eta^{\rm val}=0,1$ denotes the occupation number of the
valence quark and anti-quark states. Obviously $\A_{\rm val}^{(2)}$
also contains terms of odd powers in $\omega$. As for the contribution
of the Dirac sea (\ref{afquad}) these terms correspond to the imaginary
part in Euclidean space. Terms of odd power in $\omega$ have the
important property of removing the degeneracy between solutions with
$\pm\omega$. In the Skyrme model these terms originate from the
Wess--Zumino action \cite{wi83,ca88} which results in leading order from
the gradient expansion of the imaginary part of the quark loop
\cite{eb86}.

Finally we need to expand the purely meson part of the
action (\ref{ames})
\be
\A_m^{(2)}& = & -\frac{1}{2}m_\pi^2f_\pi^2\int d^3r
\int^{+\infty}_{-\infty}\frac{d\omega}{2\pi}
\Big\{{\rm cos}\Theta\ \tilde{\mbox{\boldmath $\eta$}}(\omega)
\cdot\tilde{\mbox{\boldmath $\eta$}}(-\omega)
\nonumber \\*
&&\qquad\qquad\qquad
+\frac{1}{4}\Big(1+\frac{m_s}{m}\Big)
\Big({\rm cos}\Theta\ +\frac{m_s^0}{m^0}\Big)
\sum_{\alpha=4}^7\tilde\eta_\alpha(\omega)
\tilde\eta_\alpha(-\omega)\Big\}
\label{amquad}
\ee
wherein we made use of the relation $G_{\rm NJL}= m^0 M/m_\pi^2f_\pi^2$,
{\it c.f.} eqns.\ (\ref{prop} - \ref{BSeq}).

Due to the invariance of the soliton configuration under grand
spin transformations the eigenvalues $\epsilon_\mu$ of (\ref{diagham})
are degenerate with respect to the corresponding projection quantum
number. It is therefore useful to decompose the fluctuations in
terms of their grand spin and parity eigenvalues. Since the unitary
matrix ${\cal T}$ in eqn. (\ref{deft}) has vanishing grand spin
this decomposition transfers directly to $\tilde h_{(1)}(\omega)$:
\be
\tilde h_{(1)}(\omega)=\sum_{GM\pi}\tilde h_{(1)}^{GM\pi}(\omega).
\label{gdec1}
\ee
Here $G$ denotes the grand spin quantum number and $M$ the
associated projection while $\pi=(-1)^G,(-1)^{G+1}$ refers
to parity. We need to evaluate expressions of the type
\be
\sum_{\mu\nu}{\cal R}(\omega;\epsilon_\mu,\epsilon_\nu)
\Big|\langle\mu|\tilde h_{(1)}(\omega)|\nu\rangle\Big|^2
\label{gdec2}
\ee
where ${\cal R}$ denotes the regulator function. As mentioned above
$\epsilon_\mu$ does not depend on the grand spin projection $M_\mu$
of the eigenstate $|\mu\rangle$. According to the Wigner--Eckart
theorem we may therefore rewrite expression (\ref{gdec2})
\be
&& \hspace{-1cm}
\sum_{n_\mu n_\nu}\sum_{G_\mu G_\nu \atop \pi_\mu \pi_\nu}
{\cal R}\left(\omega;\epsilon_\mu,\epsilon_\nu\right)
\sum_{G G^\prime \atop \pi \pi^\prime}
\langle n_\mu G_\mu \pi_\mu||\tilde h_{(1)}^{G\pi}(\omega)||
n_\nu G_\nu \pi_\nu\rangle\langle n_\mu G_\mu \pi_\mu||
\tilde h_{(1)}^{G^\prime\pi^\prime}(\omega)||
n_\nu G_\nu \pi_\nu\rangle^*
\nonumber \\ && \hspace{2cm}\times
\sum_{M_\mu M_\nu \atop M M^\prime}
\pmatrix{G_\mu & G & G_\nu \cr - M_\mu & M & M_\nu}
\pmatrix{G_\mu & G^\prime & G_\nu \cr - M_\mu & M^\prime & M_\nu}.
\label{wigeck}
\ee
Here $n_\mu$ represents quantum numbers not related to grand
spin or parity. The sum over the projection quantum numbers may be
carried out yielding $\delta_{GG^\prime}$. Parity invariance
furthermore implies $\pi=\pi^\prime$. We then get
\be
\sum_{n_\mu n_\nu}\sum_{G_\mu G_\nu \atop \pi_\mu \pi_\nu}
{\cal R}\left(\omega;\epsilon_\mu,\epsilon_\nu\right)
\sum_{G\pi} \Big|\langle n_\mu G_\mu \pi_\mu
||\tilde h_{(1)}^{G\pi}(\omega)||
n_\nu G_\nu \pi_\nu\rangle\Big|^2.
\label{gdec3}
\ee
Thus the fluctuations decouple with respect to parity and
grand spin. As in the Skyrme model this feature is inherited
from the symmetry of the soliton configuration. Furthermore
this decoupling  permits a (grand spin) partial wave analysis.
We will take advantage of this fact when considering the kaon
bound state in the next section.

\vfill\eject

\stepcounter{chapter}
\leftline{\large\it 3. The Kaon Bound State}

\medskip

The central issue of this paper is to describe hyperons as
kaon bound states in the soliton background in the spirit
of Callan and Klebanov \cite{ca85,ca88}. Thus we neglect the
pion fluctuations, \ie we consider
\be
\tilde{\mbox{\boldmath $\eta$}}({\bf r},\omega )=0 \quad {\rm and} \quad
\sum_{\alpha=4}^7\tilde\eta_\alpha({\bf r},\omega )\lambda_\alpha=
\pmatrix{0 & \tilde K({\bf r},\omega)\cr
\tilde K^\dagger({\bf r},-\omega)&0\cr}
\label{stfluc}
\ee
wherein $\tilde K({\bf r},\omega)$ is a two--component isospinor.
It is quite helpful to remember that in the case of SU(3) symmetry
($m_s^0=m^0$) the zero-mode associated with the rotation of the
soliton into strange direction corresponds to
\be
K_0({\bf r})={\bf \hat r}\cdot{\mbox{\boldmath $\tau$}} U_0
\pmatrix{{\rm sin}\frac{\Theta(r)}{2} \cr 0\cr}.
\label{zeromode}
\ee
$U_0$ is hereby an arbitrary $2\times2$ space--time independent unitary
matrix fixing the isospin orientation. This zero mode can be obtained
analytically by rewriting an infinitesimal vector transformation of the
chiral soliton (\ref{chsol}) into strange direction in terms of the
fluctuation (\ref{stfluc}). Note that this zero mode carries grand spin
$G=1/2$ and orbital angular momentum $L=1$, {\it i.e.} it represents
a kaon P--wave. Abandoning SU(3) symmetry and switching on the
strange current mass, we expect the kaon bound state to emerge in the
zero mode channel. This naturally leads to the {\it ansatz}
\be
\tilde K({\bf r},\omega)={\bf \hat r}\cdot{\mbox{\boldmath $\tau$}}
\Omega(\omega,r) \qquad {\rm with} \qquad
\Omega(r,\omega)=\pmatrix{a(r,\omega) \cr b(r,\omega)\cr}
\label{bound}
\ee
for the kaon bound state. $\Omega(r,\omega)$ is a two--component
isospinor which only depends on  the radial coordinate $r$ and the
frequency $\omega$. The angular dependence is completely given by
the P--wave ${\bf \hat r}\cdot{\mbox{\boldmath $\tau$}}$. At this
point we use that the fluctuations decouple with respect to grand
spin and parity. We may thus confine ourselves to the bound state
channel. Together with the {\it ans\"atze} (\ref{stfluc}) and
(\ref{bound}) the Hamiltonians then simplify a lot:
\be
\!\!\!\tilde h_{(1)}({\bf r},\omega)&=&-\frac{1}{2}
(m+m_s)\pmatrix{0 & u_0({\bf r})\Omega(r,\omega)\cr
\Omega^\dagger(r,-\omega) u_0({\bf r})&0\cr}
\label{h1str} \\
\nonumber \\*
\!\!\tilde h_{(2)}({\bf r},\omega,-\omega)
&=&\frac{1}{4}(m+m_s)
\pmatrix{u_0({\bf r})\beta\Omega(r,\omega)
\Omega^\dagger(r,\omega)u_0({\bf r}) & 0 \cr
0&-\beta\Omega^\dagger(r,-\omega)\Omega(r,-\omega)\cr}
\label{h2str}
\ee
For convenience we have introduced the unitary, self-adjoint
transformation matrix $u_0$
\be
u_0({\bf r})=\beta\left({\rm sin}\frac{\Theta}{2}
-i\gamma_5{\bf \hat r}\cdot{\mbox{\boldmath $\tau$}}
{\rm cos}\frac{\Theta}{2}\right).
\label{u0}
\ee
$\tilde h_{(1)}(\bf r,\omega)$ has grand spin $G=1/2$ while
$\tilde h_{(2)}(\bf r,\omega,-\omega)$ is a linear combination
of $G=0$ and $G=1$ terms. In order to evaluate ${\cal A}_F^{(2)}$
the $G=1$ piece may be dropped due to isospin invariance. With these
expressions the resulting second order action may generically be
written as
\be
\A^{(2)}[\Omega]&=&
\int_{-\infty}^{+\infty}\frac{d\omega}{2\pi}\Big\{\int drr^2
\int dr^\prime r^{\prime2}\ \Phi^{(2)}(\omega;r,r^\prime)
\Omega^\dagger(r,\omega)\Omega(r^\prime,\omega)
\nonumber \\*
&&\qquad\qquad\qquad
+\int dr r^2\ \Phi^{(1)}(r)\Omega^\dagger(r,\omega)
\Omega(r,\omega)\Big\}
\label{kernel}
\ee
A further consequence of isospin invariance is that these kernels
are unit matrices in isospace. The local kernel $\Phi^{(1)}(r)$
does not depend on the eigenenergy $\omega$. It acquires contributions
form the meson part of the action as well as those terms involving
$h_{(2)}$
\be
&&\hspace{-2cm}\Phi^{(1)}(r)=
-\frac{\pi}{2}m_\pi^2f_\pi^2\left(1+\frac{m}{m_s}\right)
\left({\rm cos}\Theta+\frac{m^0}{m_s^0}\right)
\nonumber \\ &&
-\frac{N_C}{4}\eta_{\rm val}(m+m_s)
\int \frac{d\Omega}{4\pi}\psi_{\rm val}^\dagger({\bf r})
u_0({\bf r})\beta u_0({\bf r}) \psi_{\rm val}({\bf r})
\label{phi1} \\ &&
-\frac{N_C}{4}(m+m_s) \int_{1/\Lambda^2}^\infty
\frac{ds}{\sqrt{4\pi s}}\int \frac{d\Omega}{4\pi}
\Big\{\sum_{\mu=ns}\epsilon_\mu e^{-s\epsilon_\mu^2}
\psi_\mu^\dagger({\bf r})u_0({\bf r})\beta u_0({\bf r})
\psi_\mu({\bf r})
\nonumber \\ && \hspace{7cm}
+\sum_{\rho=s}\epsilon_\rho e^{-s\epsilon_\rho^2}
\psi_\rho^\dagger({\bf r})\beta
\psi_\rho({\bf r})\Big\}.
\nonumber
\ee
The integral $\int (d\Omega/4\pi)$ indicates that we have
taken the average with regard to the internal degrees of freedom.
The bilocal kernel $\Phi^{(2)}(\omega;r,r^\prime)$ originates from the
terms quadratic in $h_{(1)}$ and is symmetric in $r$ and $r^\prime$.
\be
&&\hspace{-1cm}
\Phi^{(2)}(\omega;r,r^\prime)=
-\frac{N_C}{4}(m+m_s)^2 \int \frac{d\Omega}{4\pi}
\int \frac{d\Omega^\prime}{4\pi} \Bigg\{\eta_{\rm val} \sum_{\rho=s}
\frac{\psi_{\rm val}^\dagger({\bf r})u_0({\bf r})\psi_\rho({\bf r})
\psi_\rho^\dagger({\bf r}^\prime)u_0({\bf r}^\prime)
\psi_{\rm val}({\bf r}^\prime)}
{\epsilon_{\rm val}-\omega-\epsilon_\rho}
\nonumber \\ && \hspace{3cm}
-\sum_{\mu=ns\atop\rho=s}
\psi_\mu^\dagger({\bf r})u_0({\bf r})\psi_\rho({\bf r})
\psi_\rho^\dagger({\bf r}^\prime)u_0({\bf r}^\prime)
\psi_\mu({\bf r}^\prime){\cal R}_{\mu,\rho}(\omega)\Bigg\}
\label{phi2}
\ee
The regulator function appearing eqn. (\ref{phi2}) may be read off
from eqn. (\ref{afquad})
\be
&&\hspace{-2.5cm}{\cal R}_{\mu,\rho}(\omega)=
\int_{1/\Lambda^2}^\infty ds \sqrt{\frac{s}{\pi}}
\Big\{\frac{{\rm e}^{-s\epsilon_\mu^2}
+{\rm e}^{-s\epsilon_\rho^2}}{s}
+[\omega^2-(\epsilon_\mu+\epsilon_\rho)^2]
R_0(s;\omega,\epsilon_\mu,\epsilon_\rho)
\nonumber \\ &&\hspace{2.5cm}
-2\omega\epsilon_\rho R_1(s;\omega,\epsilon_\mu,\epsilon_\rho)
+2\omega\epsilon_\mu R_1(s;\omega,\epsilon_\rho,\epsilon_\mu)
\Big\}.
\label{phi2reg}
\ee
Note that we have decomposed the sums over the eigenstates of the
one-particle Hamiltonian (\ref{diagham}) into sums over non-strange
and purely strange states $|\mu\rangle$ and $|\rho\rangle$,
respectively. The reader may consult appendix C for more
details on these matrix elements. Varying (\ref{kernel}) with
respect to $\Omega$ yields a homogeneous linear integral equation
\be
r^2\left\{\int dr^\prime r^{\prime2} \Phi^{(2)}(\omega;r,r^\prime)
\Omega(r^\prime,\omega)+\Phi^{(1)}(\omega;r)\Omega(r,\omega)
\right\}=0
\label{eqmfluc}
\ee
which in fact is the Bethe--Salpeter equation for the kaon bound state
in the soliton background. It is the analog of the bound state
equation in the Callan--Klebanov approach \cite{ca85,ca88} to the
Skyrme model. In the present model we are even able to
give an expression for the strange valence quark wave--function
$\delta \Psi_{\rm val}^s$ which is induced by the solution of
the Bethe--Salpeter equation (\ref{eqmfluc})
\be
\delta \Psi_{\rm val}^s({\bf r},t)
&=&\int_{-\infty}^{+\infty}\frac{d\omega}{2\pi}
\delta\tilde\Psi_{\rm val}^s({\bf r},\omega)\
{\rm e}^{-i(\epsilon_{\rm val}-\omega)t}.
\label{psift}
\ee
Performing a perturbation expansion in powers of the kaon
wave--function the leading term reads
\be
\delta\tilde\Psi_{\rm val}^s({\bf r},\omega)&=&
-\frac{1}{2}(m+m_s)
\left(\epsilon_{\rm val}-\omega-h_{(0)}({\bf r})\right)^{-1}
\Omega^\dagger(r,\omega)u_0({\bf r})
\Psi_{\rm val}^{\rm ns}({\bf r}).
\label{psistr}
\ee
It can easily be verified that this strange valence (anti-)
quark carries total angular momentum $J=1/2$ and isospin $I=0$.

As usual the normalization of the bound state wave function remains
undetermined by the Bethe--Salpeter equation. For a consistent
normalization we need to calculate the strangeness of the
kaon fluctuations using the same arbitrary units as in the
Bethe--Salpeter equation. The derivation of this quantity
is described in Appendix A:
\be
 S  & = & S^{\rm val} + S^{\rm vac}
\label{str}
\\*
S^{\rm val} & = &
-\eta_{\rm val} \int\frac{d\omega}{2\pi} \sum_{\rho=s} \frac{
\langle{\rm val}|\tilde h_{(1)}({\bf r},\omega)|\rho\rangle
\langle\rho|\tilde h_{(1)}({\bf r},-\omega)|{\rm val}\rangle}
{(\epsilon_{\rm val}-\omega-\epsilon_\rho )^2}
\label{strval}
\\*
S^{\rm vac} & = & \int\frac{d\omega}{2\pi}
\int_{1/\Lambda^2}^\infty \frac{ds}{\sqrt{4\pi s}}
\sum_{\rho=s} e^{-s\epsilon_\rho^2}
\Big\{ (1-2s\epsilon_\rho^2) \langle\rho
|\tilde h_{(2)}({\bf r},\omega,-\omega)|\rho\rangle
\nonumber \\* &&\hspace{6cm}
-s\epsilon_\rho \langle\rho|\tilde h_{(1)}({\bf r},\omega)
\tilde h_{(1)}({\bf r},-\omega)|\rho\rangle \Big\}
\label{strvac}
\\*
&& - \int\frac{d\omega}{2\pi}
\int_{1/\Lambda^2}^\infty ds \sqrt{\frac{s}{4\pi}}
\sum _{\mu=ns\atop\rho=s}
\langle\mu|\tilde h_{(1)}({\bf r},\omega)|\rho\rangle
\langle\rho|\tilde h_{(1)}({\bf r},-\omega)|\mu\rangle
(\epsilon_\mu + \epsilon_\rho - \omega)
\nonumber \\* && \hspace{-1cm} \times
\Big\{R_0(s; \omega , \epsilon_\mu, \epsilon_\rho)
-s(\omega+\epsilon_\mu + \epsilon_\rho)
\Big((\epsilon_\rho-\omega)
R_1(s;\omega ,\epsilon_\mu,\epsilon_\rho)
+\omega R_2(s; \omega , \epsilon_\mu,\epsilon_\rho)
\Big)\Big\}.
\nonumber
\ee
According to refs. \cite{ca85,ca88} we introduce Fourier
amplitudes $a_i(\omega)$ for the kaon modes (\ref{bound})
of energy $\omega$
\be
a(r,\omega)=\eta_\omega(r)a_1(\omega)
\qquad {\rm and}\qquad
b(r,\omega)=\eta_\omega(r)a_2(\omega)
\label{annihil}
\ee
where the Fourier analysis in eqn. (\ref{feta}) allows for positive
as well as negative energies. In the course of quantization these
amplitudes acquire the status of annihilation operators. The radial
function $\eta_\omega(r)$ represents a solution of the Bethe--Salpeter
equation (\ref{eqmfluc}). Assuming normal order prescription we obtain
for the strangeness operator
\be
S = \int \frac{d\omega}{2\pi} \int dr \int dr^\prime
\Phi_S(\omega;r,r^\prime)\eta^*_\omega(r)\eta_\omega(r^\prime)
\left(a_1^\dagger(\omega)a_1(\omega)+
a_2^\dagger(\omega)a_2(\omega)\right).
\label{soper}
\ee
The bilocal function $\Phi_S(\omega;r,r^\prime)$ can be read off
from eqns. (\ref{str}--\ref{strvac}) together with
matrix elements listed in appendix C. Here it is only important to
note that $\Phi_S(\omega;r,r^\prime)$ is symmetric under the
exchange $r\leftrightarrow r^\prime$, implying that the integral
is real--valued. Actually $\Phi_S(\omega;r,r^\prime)$ is of the
form of $\Phi^{(2)}(\omega;r,r^\prime)$ with the appropriate
regulator function to be substituted in eqn. (\ref{phi2}).
The radial function $\eta_\omega(r)$ is then normalized
such that each occupation of the bound state yields unit strangeness
charge, {\it i.e.} we demand
\be
\Big| \int dr \int dr^\prime \Phi_S(\omega;r,r^\prime)
\eta^*_\omega(r)\eta_\omega(r^\prime) \Big| = 1 .
\label{normal}
\ee
Obviously the sign of the expectation value $\langle S \rangle$
is determined by the dynamics. From eqn. (\ref{strval}) we see that
a valence quark dominated configuration with $\epsilon_{\rm val} -
\omega>0\quad (\eta_{\rm val}=1)$ has negative strangeness. We will see later
on from our numerical studies that the solution to the Bethe--Salpeter
equation with the smallest absolute value $\omega$ always carries
negative $\langle S \rangle$.

\vfill\eject

\stepcounter{chapter}
\leftline{\large\it 4. Quantization of Spin and Isospin}
\medskip

The system consisting of the  chiral soliton and the kaon
bound state - as it stands - does not describe physical baryons
although it carries unit baryon number. It actually corresponds
to a superposition of states with arbitrary spin and isospin. In
order to project onto states with good spin and isospin we adopt
the semiclassical quantization which is commonly applied to the chiral
soliton \cite{ad83}.

This quantization procedure requires the introduction of
time dependent collective coordinates describing the
orientation in isospace. These coordinates are most
conveniently parametrized in terms of an $SU(2)$ rotation
matrix $R(t)$. Then the {\it ansatz} for our  meson fields
reads
\be
M=R(t)\xi_0\xi_f\langle\Sigma\rangle\xi_f\xi_0R^\dagger(t)
\qquad {\rm with}\qquad R(t)\in SU(2).
\label{crank}
\ee
Owing to the isosinglet character of $\langle\Sigma\rangle$
the {\it ansatz} (\ref{crank}) is equivalent to the substitution
\be
\xi_0\rightarrow R(t)\xi_0 R^\dagger(t) \qquad
{\rm and}\qquad \tilde K\rightarrow R(t)\tilde K.
\label{crank1}
\ee
Obviously this is identical to the introduction of collective
coordinates by Callan and Klebanov \cite{ca85} although the
{\it ansatz} (\ref{crank}) is different from their parametrization
of the pseudoscalar fields. This identification especially implies
that the total isospin is carried by the collective coordinates
$R(t)$ while the kaon field $\tilde K$ has lost its isospin. As
the NJL model is originally formulated in terms of quark fields this
is intuitively clear since this ``transmutation" of isospin is a
direct consequence of the fact that the strange quarks (\ref{psistr})
have zero isospin. Moreover, eqn. (\ref{crank1}) shows that
due to the hedgehog structure of $\xi_0$ the absolute value of
the spin carried by the soliton is identical to the total isospin.

The time dependence of the collective rotations is measured by
the angular velocities ${\mbox {\boldmath $\Omega$}}$
\be
\frac{i}{2}{\mbox {\boldmath $\tau$}}\cdot
{\mbox {\boldmath $\Omega$}}=R^\dagger(t){\dot R}(t).
\label{velocity}
\ee
The main goal of this section is to expand the action in powers of
{\mbox {\boldmath $\Omega$}} in the presence of the kaon bound state.
For this calculation it is convenient to switch to the flavor rotating
frame $q({\bf r},t)=R(t) q^\prime({\bf r},t)$. In this
frame the rotation matrices are eliminated from the ansatz (\ref{crank}),
however, the time derivative supplements the Dirac operator by the
Coriolis term, {\it i.e.} the Dirac operator becomes (see also
ref. \cite{re89})
\be
D^\prime_E=-\partial_\tau-h_{(0)}+\frac{i}{2}
{\mbox{\boldmath $\tau$}}\cdot{\mbox{\boldmath $\Omega$}}_E
-h_{(1)}-h_{(2)}
\label{flrot}
\ee
in Euclidean space. ${\mbox{\boldmath $\Omega$}}_E =
i{\mbox{\boldmath $\Omega$}}$ denotes the continuation of the
angular velocity into Euclidean space. Note that
${\mbox{\boldmath $\Omega$}}_E$ has to be considered a
real quantity. We then perform a perturbation expansion
in ${\mbox{\boldmath $\Omega$}}_E$ as well as in the kaon
fluctuations. The resulting expression is finally continued
back into Minkowski space. The term quadratic in
${\mbox{\boldmath $\Omega$}}$ yields the moment of
inertia $\alpha^2$ for iso-rotations and has already been
discussed in detail previously \cite{re89,go91,we92}.
Our special attention, however,
concerns the coupling of the kaon fluctuations to the
angular velocities which will determine the mass splittings
of baryons with different isospin but equal strangeness.
The lowest order term which contains this coupling and
represents an isosinglet is linear in
${\mbox{\boldmath $\Omega$}}$ and quadratic in the
amplitudes (operators) $a_i(\omega)$. Thus the contribution
to the collective Lagrangian due to the rotations reads
\be
L_{\mbox{\boldmath $\Omega$}}=
\frac{1}{2}\alpha^2{\mbox{\boldmath $\Omega$}}^2
-\frac{1}{2} \int \frac{d\omega}{2\pi} c(\omega)
{\mbox{\boldmath $\Omega$}}\cdot
\left(\sum_{i,j=1}^2 a_i^\dagger(\omega)
{\mbox{\boldmath $\tau$}}_{ij}
a_j(\omega)\right)+\cdot\cdot\cdot.
\label{lcoll}
\ee
For the moment of inertia $\alpha^2$ we adopt previous results
(\cite{re89,we93}) while the calculation for $c(\omega)$ is presented
in appendix B. Here we just quote the final result.
\be
c(\omega) & = & c_{\rm val}(\omega) + c_{\rm vac}(\omega)
\nonumber \\
c_{\rm val}(\omega) & = & \eta_{\rm val} \sum_{\mu=ns}\Big[
\frac{\langle{\rm val}|\tilde h_{(2)}(\omega,-\omega)|
\mu\rangle\langle\mu|\tau_3|{\rm val}\rangle}
{\epsilon_{\rm val}-\epsilon_\mu}\ + \ {\rm h.\ c.}\Big]
\nonumber \\ && \qquad
+\eta_{\rm val} \sum_{\mu=ns \atop \rho=s}\Big[\frac{
\langle{\rm val}|\tilde h_{(1)}(\omega)|\rho\rangle\langle\rho
|\tilde h_{(1)}(-\omega)|\mu\rangle\langle\mu|\tau_3
|{\rm val}\rangle} {(\epsilon_{\rm val}-\epsilon_\mu)
(\epsilon_{\rm val}-\omega-\epsilon_\rho)}
\ + \ {\rm h.\ c.}\Big]
\nonumber \\
c_{\rm vac}(\omega) & = &
-\int_{1/\Lambda^2}^\infty \frac{ds}{\sqrt{4\pi s}}
\Bigg\{\sum_{\mu,\nu=n.s.}\langle\mu|\tau_3|\nu\rangle
\langle\nu|\tilde h_{(2)}(\omega,-\omega)|\mu\rangle
\frac{\epsilon_\mu e^{-s\epsilon_\mu^2}
-\epsilon_\nu e^{-s\epsilon_\nu^2}}{\epsilon_\mu-\epsilon_\nu}
\label{cpar} \\ && \hspace{-0.5cm}
+\sum_{\mu,\nu=n.s. \atop \rho=s.}
\langle\mu|\tau_3|\nu\rangle\langle\nu|\tilde h_{(1)}(\omega)|
\rho\rangle\langle\rho|\tilde h_{(1)}(-\omega)|\mu\rangle
\Big[\epsilon_\nu\frac{e^{-s\epsilon_\mu^2}-e^{-s\epsilon_\nu^2}}
{\epsilon_\mu^2-\epsilon_\nu^2}+
s{\cal R}_{\mu,\nu,\rho}(\omega)\Big]\Bigg\}.
\nonumber
\ee
The somewhat complicated Feynman parameter integral
${\cal R}_{\mu,\nu,\rho}(\omega)$ which describes the quark
loops is displayed in eqn. (\ref{af12r}). The matrix elements
$\langle\nu|\tilde h_{(1)}(\omega)|\rho\rangle$ etc. are actually
evaluated by replacing $a(r,\omega)$ in the relevant expressions
in appendix C with $\eta_\omega(r)$ which is normalized according to
eqn. (\ref{normal}). Owing to eqn. (\ref{crank1}) and the hedgehog
structure of the soliton the momentum conjugate to
${\mbox {\boldmath $\Omega$}}$ is the spin carried by the soliton
\be
{\bf J}_\Theta=\frac{\partial L_{\mbox {\boldmath $\Omega$}}}
{\partial {\mbox {\boldmath $\Omega$}}}
=\alpha^2{\mbox {\boldmath $\Omega$}}
-\frac{1}{2} \int \frac{d\omega}{2\pi} c(\omega)
\left(\sum_{i,j=1}^2 a_i^\dagger(\omega)
{\mbox{\boldmath $\tau$}}_{ij}
a_j(\omega)\right)+\cdot\cdot\cdot.
\label{jtheta}
\ee
In order to extract the spin carried by the strange
fluctuations we next consider the total spin which by definition
is the expectation value
\be
\langle{\bf J}\rangle = \int D\bar q Dq \int d^3r \ q^\dagger
{\bf J} q \ {\rm exp}\left(i A_{\rm NJL}\right)
\label{spin1}
\ee
where $A_{\rm NJL}$ denotes the action associated with
the NJL Lagragian (\ref{NJL}). Since the spin operator
commutes with the collective rotations $R(t)$ we can
straightforwardly transform to the rotating frame
\be
\langle{\bf J}\rangle = \int D\bar q^\prime Dq^\prime \int d^3r
\ q^{\prime\dagger } {\bf J} q^\prime \ {\rm exp}
\left(i A^\prime_{\rm NJL}\right).
\label{spin2}
\ee
Here $A^\prime_{\rm NJL}$ represents the NJL action in the
rotating frame which also contains the Coriolis term
\be
A^\prime_{\rm NJL}=\int d^4x \left({\cal L}
-\frac{1}{2}q^{\prime\dagger}{\mbox {\boldmath $\tau$}}\cdot
{\mbox {\boldmath $\Omega$}}q^\prime\right).
\label{spin3}
\ee
According to the definition of the grand spin $\bf G$ we
may rewrite eqn (\ref{spin2}) as
\be
\langle{\bf J}\rangle = \int D\bar q^\prime Dq^\prime \int d^3r
\ q^{\prime\dagger }\left({\bf G} -
\frac{{\mbox{\boldmath $\tau$}}}{2}\right) q^\prime \ {\rm exp}
\left(i A^\prime_{\rm NJL}\right).
\label{spin4}
\ee
In this expression we may identify the soliton contribution to the
spin ${\bf J}_\Theta$ by differentiating with respect to the
angular velocity ${\mbox{\boldmath $\Omega$}}$
\be
\langle{\bf J}\rangle = \langle {\bf G} \rangle
+ \int D \bar q ^\prime D q^\prime \
\frac{1}{T}\frac{\partial A^\prime_{\rm NJL}}
{\partial{\mbox{\boldmath $\Omega$}}} \
{\rm exp}\left(i A^\prime_{\rm NJL}\right)
=\langle {\bf G} \rangle + {\bf J}_\Theta.
\label{spin5}
\ee
The spin carried by the strange fluctuations is just the
difference of the total and the soliton spins
\be
{\bf J}_K= \langle {\bf J} \rangle -{\bf J}_\Theta
= \langle {\bf G} \rangle
\label{kaonspin}
\ee
{\it i.e.} the kaon spin is identical to the expectation
value of the grand spin. Since the fluctuations also polarize
the Dirac sea which contains states with all grand spin
eigenstates the result for $\bf J_K$ may differ from the
Skyrme model result as we will see later. The calculation
is lined out in appendix A and can be summarized as
\be
{\bf J}_K = - \frac{1}{2} \int \frac{d\omega}{2\pi} d(w)
\left(\sum_{i,j=1}^2 a_i^\dagger(\omega)
{\mbox{\boldmath $\tau$}}_{ij}
a_j(\omega)\right).
\label{jk}
\ee
The spectral function $d(\omega)$ is again composed of
contributions due to valence and vacuum parts of the action
$d(\omega)=d_{\rm val}(\omega)+d_{\rm vac}(\omega)$. We get
from eqns (\ref{j3val},\ref{j3vac}) in appendix A
\be
d_{\rm val}(\omega) & = & -2\eta_{\rm val}\sum_{\rho=s}M_\rho
\frac{\langle{\rm val}|\tilde h_{(1)}({\bf r},\omega)|\rho\rangle
\langle\rho|\tilde h_{(1)}({\bf r},-\omega)|{\rm val}\rangle}
{(\epsilon_{\rm val}-\omega-\epsilon_\rho )^2}
\nonumber \\
d_{\rm vac}(\omega)& = &
2\int_{1/\Lambda^2}^\infty \frac{ds}{\sqrt{4\pi s}}
\sum_{\mu=ns}M_\mu e^{-s\epsilon_\mu^2}
\Big\{\big(1-2s\epsilon_\mu^2\big)\
\langle\mu|\tilde h_{(2)}({\bf r},\omega,-\omega)|\mu\rangle
\nonumber \\ &&\hspace{6cm}
-s\epsilon_\mu\ \langle\mu|\tilde h_{(1)}({\bf r},\omega)
\tilde h_{(1)}({\bf r},-\omega)|\mu\rangle\Big\}
\nonumber \\ &&\hspace{-0.5cm}
-2\int_{1/\Lambda^2}^\infty ds
\sqrt{\frac{s}{4\pi}} \sum_{\mu\nu} M_\nu
\langle\mu|\tilde h_{(1)}({\bf r},\omega)|\nu\rangle
\langle\nu|\tilde h_{(1)}({\bf r},-\omega)|\mu\rangle
(\epsilon_\mu + \epsilon_\nu - \omega)
\label{dpar}\\* &&\hspace{-1.5cm}\times
\Big\{R_0(s; \omega , \epsilon_\mu , \epsilon_\nu)
-s(\omega+\epsilon_\mu + \epsilon_\nu)
\Big((\epsilon_\nu-\omega)
R_1(s;\omega ,\epsilon_\mu ,\epsilon_\nu)
+\omega R_2(s; \omega , \epsilon_\mu , \epsilon_\nu)
\Big)\Big\}.
\nonumber
\ee
Again, the radial function $\eta_\omega(r)$ is to be
substituted in $\tilde h_{(i)}$. The similarity between
the expression for the strangeness charge (\ref{str}-\ref{strvac})
and (\ref{dpar}) is obvious. The appearance of the grand spin projection
$M_\nu$ for the intermediate states $|\nu\rangle$ in (\ref{dpar}) remarks
the main difference. Actually, if one confines oneself to grand spin
zero states in the sum over the non-strange states $|\mu\rangle$
in both expressions one obtains $d(\omega)=1$. Especially, taking
into account only the valence quarks yields a unit value for $d(\omega)$.
This is exactly the result of Callan and Klebanov \cite{ca85,ca88}.
However, since the evaluation of the fermion determinant involves
the whole tower of grand spin states a correction to $d(\omega)=1$
arises in the NJL model. We would furthermore like to mention that
$d_{\rm vac}$ in eqn. (\ref{dpar}) may also be obtained from
$c_{\rm vac}$ in  eqn. (\ref{cpar}) by replacing $\tau_3$ with
$G_3$ and taking into account that the eigenstates $|\mu\rangle$
of $h_{(0)}$ are also grand spin eigenstates.

The ongoing discussion will be confined to the bound state
solution of the Bethe--Salpeter equation (\ref{eqmfluc}) with
the energy eigenvalue $|\omega_0|<m_K$. Stated otherwise, we will
consider only those baryon states which are constructed by the
occupation of this bound state. It is then useful to define
\be
c(\omega_0)=c \qquad {\rm and} \qquad d(\omega_0)=d
\label{defpara}
\ee
since the annihilation and creation operators $a_i(\omega)$ project
out these quantities form the spectral integrals. With this
restriction the collective Hamiltonian obtained from
$L_{\mbox{\boldmath $\Omega$}}$ reads
\be
H_{\mbox{\boldmath $\Omega$}}=
\frac{1}{2\alpha^2}\left({\bf J}_\Theta +
\frac{c}{2}\sum_{i,j=1}^2 a_i^\dagger
{\mbox{\boldmath $\tau$}}_{ij} a_j\right)^2
=\frac{1}{2\alpha^2}\left({\bf J}_\Theta
+\chi{\bf J}_K\right)^2
\label{homega}
\ee
where we have introduced the parameter $\chi=-c/d$ and omitted
the arguments of the annihilation (creation) operators. Obviously
(\ref{homega}) is identical to the Callan--Klebanov result \cite{ca88}
with $c$ substituted by $\chi$. Since the total spin is the sum
${\bf J}= {\bf J}_\Theta + {\bf J}_K$ and the absolute value
of ${\bf J}_\Theta$ equals the total isospin, as discussed after
eqn. (\ref{crank1}) $H_{\mbox{\boldmath $\Omega$}}$ becomes
\be
H_{\mbox{\boldmath $\Omega$}}=
\frac{1}{2\alpha^2}\left(\chi J(J+1) + (1-\chi)I(I+1)
+(\frac{d\chi}{2})^2 S(S-2)\right).
\label{homega1}
\ee
Here $J$ and $I$ denote the spin and isospin of the baryon under
consideration. We have furthermore used that
$\left(\sum_{i,j=1}^2 a_i^\dagger
{\mbox{\boldmath $\tau$}}_{ij} a_j\right)^2$ may be written
as $S(S-2)$. This term is already of forth order in the
kaon fluctuations and should thus been dropped for
consistency\footnote{In the SU(3) symmetric case the nucleon is
degenerate with the hyperons with the same spin. Since we then have
$\chi=1$ there additional terms of forth must occur in order to
cancel the last term in (\ref{homega1}). As all forth order terms
have been ignored this provides an additional argument to neglect
this expression.} In order to finally reach the mass formula we
note that each occupation of the kaon bound state corresponds
to replacing a non-strange valence quark by a strange valence
quark. According to eqn. (\ref{psift}) this shifts the total
energy by $S\omega_0$ since we have $S\le0$ for physical
baryons. We may now conclude this section by presenting
the mass formula for strange and non-strange baryons
\be
M_B=E_{\rm cl}+S\omega_0+\frac{1}{2\alpha^2}
\left(\chi J(J+1)+(1-\chi)I(I+1)\right)
\label{mass}
\ee
wherein $E_{\rm cl}$ is the classical energy due to the static
soliton configuration. We would finally like to note that the
baryon mass formula (\ref{mass}) actually does not depend on the
normalization of the strange bound state since $\chi=-c/d$ is the
ratio of two objects quadratic in the bound state wave--function
$\eta_{\omega_0}(r)$.
\vskip 1.5cm

\stepcounter{chapter}
\leftline{\large\it 5. Numerical Results and Discussion}

\medskip

The numerical investigation is performed in several steps. First, the
self--consistent chiral soliton which minimizes the static
energy functional is constructed. In the second step this solution
is employed to evaluate the kernels $\Phi^{(i)}$ of the
Bethe--Salpeter equation (\ref{eqmfluc}) for a given kaon
frequency $\omega$. The descretized version of the
Bethe--Salpeter equation is extended to an eigenvalue
problem
\be
r_i^2\sum_{j}\left\{\Delta r\ r_j^2\ \Phi^{(2)}_{ij}(\omega)
+\delta_{ij}\Phi^{(1)}_{j}\right\}\eta_\omega(j)
=\lambda(\omega)\eta_\omega(i)
\label{eigprob}
\ee
wherein $i$ and $j$ label the lattice points $r_i=i\Delta r$.
Obviously the eigenvalues $\lambda$ depend on the
frequency $\omega$. The actual solution to the Bethe--Salpeter
equation corresponds to $\lambda(\omega)=0$, {\it i.e.} we
adjust $\omega_0$ such that eqn. (\ref{eigprob}) has a
vanishing eigenvalue.

\begin{figure}
\centerline{\hskip -1.5cm
\psfig{figure=bound.tex.ps,height=9.0cm,width=16.0cm}}
\fcaption{The bound state energy $\omega_0$ as a function of the
constituent quark mass $m$.}
\end{figure}

In figure (5.1) we display this solution
$\omega_0$ of as a function of the constituent quark mass
$m$. It is remarkable that $\omega_0$ is a continuous function
of $m$; especially since $\epsilon_{\rm val}$ changes sign
at $m\approx760$ MeV and thus $\eta_{\rm val}=0$ for constituent
quark masses larger than this. This feature, of course, represents
an excellent check on the analytical as well as numerical calculations.
It should be noted that the Feynman parameter integrals
eqn. (\ref{regfct}) need to be treated very carefully in case a
small energy eigenvalue $\epsilon_\mu$ is involved. {\it I.e.} in
this case numerically stable results can only be achieved when
the integrals are evaluated using a very dense grid.

In the third step the resulting wave--function $\eta_{\omega_0}$
is normalized according to eqn. (\ref{normal}). In figure (5.2) we
display these solutions for various strengths of the symmetry breaking
which is manifested in different kaon masses $m_K$. We observe a
distinct localization for the physical value of the kaon mass
$m_K=495$MeV. We also see that for the symmetric case
$m_K=m_\pi=135$MeV the zero mode wave--function $N_0{\rm sin}\Theta/2$
is perfectly reproduced, a further check on our numerical routines.
The normalization $N_0$ is chosen to fulfill the condition (\ref{normal})
with $\eta(r)=N_0{\rm sin}\Theta/2$ and $\Phi_S(\omega=0;r,r^\prime)$.
We should note that the solution in the symmetric case does not appear
exactly at $\omega_0=0$, we rather find $\omega_0=-0.2$MeV\footnote{In
ref. \cite{we93} we found -1.4MeV. Fortunately we were able to improve
the numerical code yielding a significantly smaller numerical error.}.
This result indicates that the numerical error of the bound state energy
is as small as 0.2MeV.

\begin{figure}
\centerline{\hskip -1.5cm
\psfig{figure=bzero.tex.ps,height=9.0cm,width=16.0cm}}
\fcaption{The bound state wave--function $\eta_{\omega_0}(r)$
as a function of radial distance $r$ for various strengths of
the symmetry breaking. Also shown is the zero mode
wave--function $N_0{\rm sin}\Theta/2$. In all cases the
constituent quark mass is $m=$400MeV.}
\end{figure}

In figure (5.3) the valence and vacuum contributions to the
strangeness are resolved. We see that the strangeness
charge is valence quark dominated as long as the valence
quark state has not joined the Dirac sea. We should stress
that the normalization integral stays negative even for
$\eta_{\rm val}=0$, {\it i.e.} the kaon bound state always has
$S=-1$ and the strangeness charge is continuous at $m\approx760$MeV.
Due to the normalization (\ref{normal}) the fact that the strangeness
charge is a smooth function of $m$ thus is manifested by the bound
state wave--function being continuous when the valence quark joins
the Dirac sea.  This can be observed from figure (5.4) where the
bound state wave function is plotted for various constituent quark
masses.

\begin{figure}
\centerline{\hskip -1.5cm
\psfig{figure=strangeness.tex.ps,height=9.0cm,width=16.0cm}}
\fcaption{Valence and Dirac sea contribution to the strangeness
charge associated with the kaon bound state as a function of
the constituent quark mass $m$.}
\end{figure}

\begin{figure}
\centerline{\hskip -1.5cm
\psfig{figure=kaon.tex.ps,height=9.0cm,width=16.0cm}}
\fcaption{The bound state wave--function $\eta_{\omega_0}(r)$
as a function of radial distance $r$ for three different
constituent quark masses $m$.}
\end{figure}

We are finally able to numerically calculate the parameters
$c$ and $d$ associated with the bound state wave--function
$\eta_{\omega_0}$. The results are shown in figure (5.5).
There is a notable reduction from the grand spin zero
result $d=1$. For small constituent quark masses the soliton
is valence quark dominated and thus $d$ approaches unity
in that limit. However, even when the valence quark has joined
the Dirac sea the reduction from unity is not larger than
about 15\%. Since both $c$ and $d$ decrease with increasing
$m$ the dependence of $\chi=-c/d$ on $m$ is somewhat
stressed.

\begin{figure}
\centerline{\hskip -1.5cm
\psfig{figure=cdchi.tex.ps,height=9.0cm,width=16.0cm}}
\fcaption{The parameters $c,d$ and $\chi$ defined in section 4 as
functions of the constituent quark mass $m$.}
\end{figure}

Let us now discuss the implications of our results on the spectrum
of the low--lying $\frac{1}{2}^+$ and $\frac{3}{2}^+$ baryons. Upon
inversion of the mass formula (\ref{mass}) we get some information of
the ``empirical" values of the parameters $\omega_0,\alpha^2$ and
$\chi$.  Since the nucleon and $\Xi$ baryons have both equal spin
and isospin we may extract from (\ref{mass})
\be
M_\Xi-M_n=-2\omega_0\approx379{\rm MeV}
\qquad {\rm or}\qquad \omega_0\approx-189.5{\rm MeV}
\label{omegaexpt}
\ee
which favors a constituent quark mass $m$ somewhat below 400MeV
(see also table (5.1)). On the other hand the ``empirical" value for
the moment of inertia $\alpha^2$ is given by
\be
M_\Delta-M_n=\frac{3}{2\alpha^2}\approx293{\rm MeV}
\qquad {\rm or}\qquad \alpha^2\approx5.12{\rm GeV}^{-1}
\label{alphaexpt}
\ee
which is satisfied for $m\approx430$MeV (see {\it e.g.} figure 1
of ref. \cite{go91}.). Finally we get for $\chi$ from states with
identical strangeness \cite{ca88}
\be
\frac{M_{\Sigma^*}-M_\Sigma}{M_\Sigma-M_\Lambda}=
\frac{3\chi}{2(1-\chi)}\approx2.49
\qquad {\rm or}\qquad\chi\approx0.62.
\label{chiexpt}
\ee
{}From figure (5.5) as well as table (5.1) we observe that this
value is obtained for $m\approx500$MeV. Apparently there is
no value of $m$ such that all three conditions (\ref{omegaexpt},
\ref{alphaexpt}, \ref{chiexpt}) are satisfied, However, the results
of our calculations  are reasonably close to the ``empirical" values
in the region 400MeV$<m<$450MeV.

\vfill\eject

\tcaption{Parameters for describing the Hyperon spectrum as
functions of the constituent mass $m$. Also listed are the
empirical values which are obtained by the consideration of
certain mass differences, see text.}
{}~
\newline
\centerline{
\begin{tabular}{|c|c|c|c|c|c|}
   \hline
$m$(MeV) & 350 & 400 & 450 & 500 & empir. \\
\hline
$\omega$(MeV) & -207.1 & -182.6 & -163.6 & -148.8 & -189.5 \\
\hline
$c$ & -0.20 & -0.36 & -0.46 & -0.53 & ---\\
\hline
$d$ & 0.90 & 0.89 & 0.89 & 0.89 & ---\\
\hline
$\chi=-c/d$ & 0.22 & 0.40 & 0.52 & 0.60 & 0.62 \\
\hline
$\alpha^2(1/({\rm GeV}))$ & 8.30 & 5.80 & 4.78 & 4.17 & 5.12 \\
\hline
\end{tabular}}
{}~
\newline

In order to judge our treatment of the NJL model we wish to
compare the results on the baryon spectrum with our earlier
calculations \cite{we92} in the Yabu--Ando approach \cite{ya88}
as well as the experimental data. For getting consistent treatments
we fix the constituent quark mass $m$ in both approaches by demanding
the experimental value for nucleon--$\Delta$ mass difference
$M_\Delta-M_n=293$MeV. As can be seen from eqn. (\ref{mass}) the SU(2)
rigid rotator spectrum is maintained for states with vanishing
strangeness and $I=J$. We thus get from the above condition the same
result as in the two--flavor model: $m=430$MeV\footnote{Fixing $m$ from
other mass differences involves the parameter $\chi$; {\it e.g.}
the difference of the mass centers in the $J=3/2$ and $J=1/2$
multiplets requires to fix $(3+5\chi)/16\alpha^2=231$MeV. Determining
$m$ from such a condition is obviously more involved numerically
although better agreemenent with the experimental data for the
mass differeneces might be obtained.}. For the parameters
involving the bound state wave function we find in this case:
$\omega_0=-170.6{\rm MeV},c=-0.42, d=0.89$ and $\chi=0.48$. In this
calculation $f_\pi$ has been kept at its experimental value.
Unfortunately, the kaon decay constant is predicted too small
$f_K=98.5$MeV. Contrarily, one might want to maintain the physical value
$f_K=113$MeV and simultaneously demand the nucleon--$\Delta$ mass
difference leading to $m=458$MeV. For this parameter set we find
$f_\pi=106.1$MeV. The quantities describing the baryon spectrum
are then $\omega_0=-178.4{\rm MeV},c=0.39,d=0.90$ and $\chi=0.44$. In
both cases the ratio $f_K/f_\pi$ turns out to be only 1.06 while the
experimental value is 1.22. The ratio $\Lambda/m$ which defines the scale
for the calculations in the soliton sector differs by only 7\% for these
two parameter sets ({\it cf.} table (2.1)) indicating that the results
on baryon properties will not change drastically. In table (5.2) we
present the results on the baryon mass differences for these two parameter
sets. We observe that by fitting $f_K$ the agreement with the experimental
data is slightly improved. This suggests that the too low predicted
mass differences are merely due to the incorrect prediction of the
ratio $f_K/f_\pi$ in the meson sector of the NJL model.

In the Yabu--Ando approach, on the other hand, the SU(2)
wave--functions of the $S=0$ and $I=J$ states are only recovered in
the infinitely strong symmetry breaking limit ($m_K\rightarrow\infty$)
\cite{ya88}. In the symmetric case $m_K=m_\pi$ the relation
(\ref{alphaexpt}) also holds in the Yabu--Ando treatment, however,
for the physical value $m_K=495$MeV a significant deviation occurs.
Following the calculation of ref. \cite{we92} then provides the
experimental $\Delta$--nucleon mass differences for $m=407$MeV.
The predicted kaon decay constant then is $f_K=99.8$MeV.
When $f_K$ is taken at its physical value the fit to the
$\Delta$--nucleon splitting requires $m=433$MeV which gives
$f_\pi=104.9$MeV. As in the Callan--Klebanov approach
the agreement of the mass differences is slightly improved
when the physical value for $f_K$ is adopted. In both approaches
the mass differences increase by up to about 10MeV when
choosing to fit $f_K$ instead of $f_\pi$.

{}~
\newline
\tcaption{The mass differences of the low-lying $\frac{1}{2}^+$
and $\frac{3}{2}^+$ baryons with respect to the nucleon. We
compare the predictions of the Callan--Klebanov (CK) and
Yabu--Ando (YA) approaches to the NJL model to the experimental data.
In both cases the up-quark constituent mass $m$ is chosen such
that the $\Delta$--nucleon mass difference is reproduced
correctly, see text.  The results for the YA approach are obtained
by employing the procedure described in ref.\cite{we92}. All data
are in MeV.}
{}~
\newline
\centerline{
\begin{tabular}{|c||c|c||c|c||c|}
\hline
& \multicolumn{2}{c||}  {CK} &
\multicolumn{2}{c||} {YA} & Expt.\\
\hline
& $f_\pi$ fixed & $f_K$ fixed
& $f_\pi$ fixed & $f_K$ fixed & \\
\hline
$m$       & 430 & 458 & 407 & 433 & --- \\
\hline
\hline
$\Lambda$ & 132 & 137 & 105 & 109 & 177 \\
\hline
$\Sigma$ & 234 & 247 & 148 & 151 & 254 \\
\hline
$\Xi$ & 341 & 357 & 236 & 243 & 379 \\
\hline
$\Delta$ & 293 & 293 & 293 & 293 & 293 \\
\hline
$\Sigma^*$ & 374 &375 & 387 & 391 & 446 \\
\hline
$\Xi^*$ & 481 & 485 & 482 & 489 & 591  \\
\hline
$\Omega$ & 613 & 622 & 576 & 586 & 733 \\
\hline
\end{tabular}}
{}~
\newline
The results displayed in table (5.2) clearly show that for the NJL
soliton model - as it stands - the Callan--Klebanov approach works
better than the treatment proposed by Yabu and Ando. Although the
difference in the two approaches is only marginal for the $\Sigma^*$
and $\Xi^*$ baryons the predictions for the $\frac{1}{2}^+$ baryons
in the Callan--Klebanov approach are much closer to the experimental
data than those of the Yabu--Ando treatment. We would like to add a
word of caution to this result since it is not obvious that the
observed superiority of the Callan--Klebanov approach can be generalized
to other models. In ref. \cite{we92} it was demonstrated that once the
parameters of the collective Hamiltonian are tuned such that they
correspond to the physical values of the decay constants the Yabu--Ando
approach yields perfect agreement between the predicted and measured
mass difference. The corresponding, very crude approximation for the
Callan--Klebanov approach consists of replacing $\omega_0$ with
$(f_K^{\rm expt} f_\pi^{\rm pred}/ f_K^{\rm pred}f_\pi^{\rm expt})
\omega_0$. This procedure then gives a somewhat too large splitting
$M_\Xi-M_n \approx400$MeV when the nucleon--$\Delta$ mass difference is
kept at is empirical value. All other mass differences involve the
parameter $\chi$ for which we do not have such a simple prescription to
extrapolate to the physical decay constants. If it turned out that
(as discussed above) the short--comings of both approaches are merely
inherited from the meson sector one could imagine that in a model which
predicts $f_\pi$ and $f_K$ correctly the baryon mass differences might
get overestimated by the Callan--Klebanov treatment.

We also would like to comment on an obvious application of the
bound state approach: the heavy quark limit. This limit provides
a further check on our results because in the heavy quark limit
the parameter $\chi$ is known to vanish \cite{gu93}. Of course,
there are conceptual problems in the NJL model with proper time
regularization once flavors are considered which possess a current
mass larger than the cut--off $\Lambda$. For the constituent up--quark
mass of 400MeV we do not encounter this problem unless the mass of the
heavy pseudoscalar meson is larger than 750MeV. In table (5.3) we display
the parameters appearing in the baryon mass formula (\ref{mass}) as a
function of the mass of the heavy meson $m_H$. The tendency
$\chi\rightarrow0$ as $m_H\rightarrow\infty$ is clearly exhibited. Of
course, the absolute value of the bound state energy $\omega_0$ rises
with $m_H$. The moment of inertia does not depend on $m_H$ and is
therefore not shown in table (5.3).

{}~
\newline
\tcaption{The parameters for the baryon mass formula (\ref{mass})
which depend on the mass of the heavy meson $m_H$. The up--quark
constituent mass is 400MeV.}
{}~
\newline
\centerline{
\begin{tabular}{|c|c|c|}
\hline
$m_H$(MeV) & $\omega_0$(MeV) & $\chi$ \\
\hline
300 & -74.4  & 0.54 \\
\hline
400 & -128.1 & 0.47  \\
\hline
500 & -185.6 & 0.40  \\
\hline
600 & -245.6 & 0.34  \\
\hline
700 & -307.6 & 0.29  \\
\hline
\end{tabular}}

\vskip 1.5cm

\stepcounter{chapter}
\leftline{\large\it 6. Conclusions}

\medskip

In this paper we have presented the description of the
low--lying $\frac{1}{2}^+$ and $\frac{3}{2}^+$ baryons in the
NJL chiral soliton model using the Callan--Klebanov or bound
state approach to strangeness. The main feature of this approach
is the existence of a P--wave kaon bound state. In the NJL model
we have determined the bound state energy and wave--function
by solving the Bethe--Salpeter equation for the K--meson in the
soliton background. Here we have devoted the main effort into the
inclusion of the collective rotations in order to quantize the
system consisting of the static soliton and the kaon bound state.
The central issue then has been to evaluate the coupling between this
bound state and the collective rotations as given by the parameter
$c$ in eqn. (\ref{cpar}). For the spin of the bound state
${\bf J}_K=-d\langle{\mbox{\boldmath $\tau$}}/2\rangle$ we have
observed a deviation from the Skyrme model result $d=1$ in the NJL
model since higher than grand spin zero quark spinors are involved
in the evaluation of the fermion determinant. These states experience
a non--vanishing polarization caused by the strange bound state.
Nevertheless, with the appropriate redefinition $\chi=-c/d$ also the
NJL model yields the Callan--Klebanov mass formula (\ref{mass}). We
have then seen that the bound state approach provides a reasonable
but not outstandingly good agreement for the predicted baryon mass
differences with the experimental data. We have also compared these
results to our earlier calculations in the Yabu--Ando treatment. For
this purpose we have imposed the same rules, {\it i.e.} reproducing the
experimental value for the $\Delta$--nucleon mass splitting, to fix
the constituent quark mass (the only free parameter) in both
approaches. This has led us to the conclusion that for the
$\frac{1}{2}^+$ baryons the bound state treatment provides a
significantly better agreement for the baryon masses measured relative
to the nucleon mass than does the collective approach. For the
$\frac{3}{2}^+$ baryons almost no difference between the results of
the two approaches has been found. Unfortunately in both approaches the
mass differences of baryons with different strangeness are predicted
too small. This obviously demands for more strength in SU(3) symmetry
breaking. We have conjectured that this problem is inherited
from the meson sector of the NJL model where in the proper--time
regularization scheme the ratio $f_K/f_\pi$ is underestimated by
about 15\%. A possible remedy could be an alternative
regularization scheme which improves on this ratio. This is even
more strongly called for when one wishes to consider heavier
quarks than the strange quark in the context of the heavy quark
symmetry. Especially since the bound state approach has been noticed
to be the appropriate tool to investigate this symmetry in the
framework of soliton models for baryons. Also the investigation of
the S--wave K--meson \cite{bl89} in the background of the NJL soliton
in order to describe the $\Lambda(1405)$ hyperon would be an
interesting challenge to pursue.
\vskip 1.5cm

\appendix

\stepcounter{chapter}
\leftline{\large\it Appendix A: Strangeness and Spin of the
Kaon Bound State}

\medskip

In this appendix we will describe the derivation of the expressions
for the strangeness (\ref{str} - \ref{strvac}) and the spin
(\ref{jk},\ref{dpar}) of the kaon bound state with bound state
energy $\omega$. We should stress that for the standard phase
convention free strange quark states carry strangeness -1.

Of course, the contribution to the strangeness due to the explicit
occupation of the valence quark level is just given as the negative
spatial integral of the associated density:
\be
S^{\rm val} & = & -\eta_{\rm val}\int d^3r
\delta\tilde\Psi_{\rm val}^{s\dagger}({\bf r},\omega)
\delta\tilde\Psi_{\rm val}^s({\bf r},\omega)
\nonumber \\
 & = & -\eta_{\rm val} \sum_{\rho=s } \frac{
\langle{\rm val}|\tilde h_{(1)}({\bf r},\omega)|\rho\rangle
\langle\rho|\tilde h_{(1)}({\bf r},-\omega)|{\rm val}\rangle}
{(\epsilon_{\rm val}-\omega-\epsilon_\rho )^2} \ .
\label{strvalapp}
\ee

According to our phase conventions the expectation value
of the strangeness charge in Minkowski space is
given by
\be
S = \int {\cal D}\bar q {\cal D} q \int d^3r
q^\dagger (-\hat S) q\  {\rm exp}(i\int d^4x \La)
\label{smink}
\ee
wherein $\La$ is the NJL - Lagrangian of eqn. (\ref{NJL}).
$\hat S  = {\rm diag}(0,0,1)$ denotes the strangeness projection
operator in flavor space. Continuation to Euclidean space
($x_0\rightarrow -ix_4 = -i\tau$) yields for this defining equation
\be
S = \frac{-i}{T}\frac{\partial}{\partial s_4}
\int {\cal D}q^\dagger {\cal D} q \exp\left\{\int d^4x_E \
(\La^{(E)}-is_4 q^\dagger \hat S q)\right\}\Big|_{s_4=0}.
\label{seucl}
\ee
Here $T$ denotes the Euclidean time interval under consideration
and  $\La^{(E)}$ is the analytic continuation of the
NJL - Lagrangian (\ref{NJL}). Application of standard bosonization
techniques \cite{eb86} to the expression (\ref{seucl}) leads to
\be
S = \frac{-i}{T}\frac{\partial}{\partial s_4}
\Tr\ {\rm log}\left(D_E(s_4)\right)\Big|_{s_4=0}
\qquad {\rm with}\qquad D_E(s_4)= -\partial_\tau - h -is_4\hat S
\label{defstr}
\ee
The Hamiltonian $h$ is defined in eqn. (\ref{de1}). We will
consider the contributions due to the real and imaginary parts
of the Euclidean action separately:
$S = S_R +S_I$ with
\be
S_{R,I}=\frac{-i}{T}\frac{\partial}{\partial s_4}
\A_{R,I}(s_4)\Big|_{s_4=0}.
\label{sreim}
\ee
Real and imaginary parts are defined via
\be
\A_R(s_4) = \frac{1}{2}\Tr\ {\rm log}\left(D_E(s_4) D_E^\dagger(s_4)\right)
\quad {\rm and} \quad
\A_I(s_4) = \frac{1}{2}\Tr\ {\rm log}\left(D_E(s_4)
\left(D_E^\dagger(s_4)\right)^{-1}\right).
\label{arai}
\ee
For the evaluation of the real part we again employ Schwinger's
proper time regularization prescription. Expanding in powers of the
real quantity $s_4$
\be
D_E(s_4) D_E^\dagger(s_4)=-\partial_\tau^2+h^2+[\partial_\tau,h]
-i\left\{\partial_\tau,s_4\hat S\right\}
-i\left[h,s_4\hat S\right]+{\cal O}(s_4^2)
\label{expandds}
\ee
we find
\be
S_R=-\frac{1}{2T}\Tr\int_{1/\Lambda^2}^\infty
ds\left(\{\partial_\tau,\hat S\}+[h,\hat S]\right)
\exp\left(\partial_\tau^2-h^2-[\partial_\tau,h]\right).
\label{sr1}
\ee
This expression is expanded in powers of meson fluctuations
({\it cf.} eqn. (\ref{de2})) by using the general formula
\be
&&\hspace{-1.5cm}
\exp\left(A+\delta A\right)= \exp\left(A\right)
+\int_0^1 d\alpha \exp\left(\alpha A\right)\delta A
\exp\left((1-\alpha)A\right)
\nonumber \\ &&\hspace{-1.0cm}
+\int_0^1 d\alpha \int_0^{1-\alpha}d\beta \exp\left(\alpha A\right)
\delta A \exp\left((1-\alpha-\beta)A\right)
\delta A \exp\left(\beta A\right)
+{\cal O}(\delta A^3).
\label{expaexpo}
\ee
Due to the cyclic property of the trace one Feynman parameter
integral can be carried out trivially at each order of $\delta A$.
Applying (\ref{expaexpo}) to eqn. (\ref{sr1}) then gives
\be
T S_R &=& \Tr\int_{1/\Lambda^2}^\infty
ds \hat S \partial_\tau \hat K_0(s)
+\Tr\int_{1/\Lambda^2}^\infty sds \hat S \partial_\tau
\left([\partial_\tau,h_{(2)}]
+\{h_{(0)},h_{(2)}\}+h_{(1)}^2\right)\hat K_0(s)
\nonumber \\ &&
-\Tr\int_{1/\Lambda^2}^\infty s^2ds \hat S \partial_\tau
\int_0^1 xdx \hat K_0(xs)\left([\partial_\tau,h_{(1)}]
+\{h_{(0)},h_{(1)}\}\right)
\nonumber \\ && \qquad\qquad\qquad\qquad\times
\hat K_0((1-x)s)\left([\partial_\tau,h_{(1)}]
+\{h_{(0)},h_{(1)}\}\right)
\label{sr2} \\ &&
+\frac{1}{2}\Tr\int_{1/\Lambda^2}^\infty sds [h_{(1)},\hat S]
\int_0^1dx \hat K_0(xs)\left([\partial_\tau,h_{(1)}]
+\{h_{(0)},h_{(1)}\}\right)\hat K_0((1-x)s)
\nonumber
\ee
where we again have made use of the zeroth-order heat kernel
$\hat K_0(s)=\exp\left(s(\partial_\tau^2-h_{(0)}^2)\right)$.
The first two terms on the $RHS$ of eqn. (\ref{sr2}) turn out to
be total time derivatives and thus vanish. The temporal part of the
trace is evaluated by introducing eigenstates $|\tau\rangle$ of the
Euclidean time. Noting that \cite{re89}
\be
\langle\tau|\hat K_0(s)|\tau^\prime\rangle=
\frac{1}{\sqrt{4\pi s}}\exp(-sh_{(0)}^2)
\exp\left(-\frac{(\tau-\tau^\prime)^2}{4s}\right)
\label{zerokernel}
\ee
this calculation amounts to carrying out Gaussian integrals.
As an example let us consider the last term in (\ref{sr2}).
We get
\be
&&\frac{1}{2} {\it Tr} \int_{1/\Lambda^2}^\infty \frac{ds}{4\pi}
\int_0^1 \frac{dx}{\sqrt{x(1-x)}} \int d\tau d\tau^\prime
\int \frac{d\omega}{2\pi}\frac{d\omega^\prime}{2\pi}
\exp\left(-\frac{(\tau-\tau^\prime)^2}{4xs}\right)
\exp\left(-\frac{(\tau-\tau^\prime)^2}{4(1-x)s}\right)
\nonumber \\ &&\times
e^{-i\omega\tau}e^{-i\omega^\prime\tau^\prime}
[\tilde h_{(1)}(i\omega),\hat S]e^{-sxh_{(0)}^2}
\left(-i\omega^\prime\tilde h_{(1)}(i\omega^\prime)+
\{h_{(0},\tilde h_{(1)}(i\omega^\prime)\}\right)
e^{-s(1-x)h_{(0)}^2}
\label{srcal1}
\ee
wherein {\it Tr} refers to the trace over spatial and
internal degrees of freedom.  We have furthermore used that
$h_{(1)}$ is local in time
\be
\langle \tau |h_{(1)}|\tau^\prime\rangle =
\int \frac{d\omega}{2\pi} \tilde h_{(1)}(i\omega)
e^{-i\omega\tau} \delta(\tau-\tau^\prime).
\label{srcal2}
\ee
Shifting the integration variable
$\tau^\prime\rightarrow \tau^\prime +\tau$ and subsequent
integration provides the desired $\delta$--function for the
energies of meson fluctuations. The remaining Gaussian integral
may readily be evaluated yielding
\be
&&\hspace{-2cm} \frac{1}{2}{\it Tr} \int_{1/\Lambda^2}^\infty ds
\sqrt{\frac{s}{4\pi}} \int_0^1 dx\int
\frac{d\omega}{2\pi} e^{-sx(1-x)\omega^2}
\nonumber \\&& \times
[\tilde h_{(1)}(i\omega),\hat S]e^{-sxh_{(0)}^2}
\left(i\omega^\prime\tilde h_{(1)}(-i\omega)+
\{h_{(0)},\tilde h_{(1)}(-i\omega)\}\right)
e^{-s(1-x)h_{(0)}^2}.
\label{srcal3}
\ee
The remaining trace is evaluated in the eigenbasis of $h_{(0)}$
(\ref{diagham}). Since $\hat S$ projects onto purely strange
states and $h_{(1)}$ changes strangeness by one unit we obtain
\be
&&\hspace{-2cm} \frac{1}{2} \int_{1/\Lambda^2}^\infty ds
\sqrt{\frac{s}{4\pi}} \int_0^1 dx\int
\frac{d\omega}{2\pi} e^{-sx(1-x)\omega^2}
\nonumber \\&& \times
\sum_{\mu={ns} \atop \rho=s}
\Big\{\langle \mu |\tilde h_{(1)}(i\omega) |\rho\rangle
\langle \rho |\tilde h_{(1)}(-i\omega) |\mu\rangle
(i\omega+\epsilon_\mu+\epsilon_\rho)
e^{-sx\epsilon_\rho^2}e^{-s(1-x)\epsilon_\mu}
\nonumber \\&& \hspace{1cm}
-\ \langle \rho |\tilde h_{(1)}(i\omega) |\mu\rangle
\langle \mu |\tilde h_{(1)}(-i\omega) |\rho\rangle
(i\omega+\epsilon_\mu+\epsilon_\rho)
e^{-sx\epsilon_\mu^2}e^{-s(1-x)\epsilon_\rho^2}\Big\}.
\label{srscal4}
\ee
Taking advantage of the symmetries under $\mu\leftrightarrow\rho$,
$x\leftrightarrow 1-x$ and $\omega \leftrightarrow -\omega$
(\ref{srscal4}) simplifies to
\be
&& \int_{1/\Lambda^2}^\infty ds \sqrt{\frac{s}{4\pi}}
\int_0^1 dx\int \frac{d\omega}{2\pi} (i\omega)
\nonumber \\ &&\hspace{2cm}\times
\sum_{\mu={ns} \atop \rho=s}
e^{-s\big[x(1-x)\omega^2+x\epsilon_\mu^2+(1-x)\epsilon_\rho^2\big]}
\langle \mu |\tilde h_{(1)}(i\omega) |\rho\rangle
\langle \rho |\tilde h_{(1)}(-i\omega) |\mu\rangle.
\label{srsacl5}
\ee
The remaining terms of (\ref{sr2}) are treated similarly under the
prospect that
\be
\langle \tau |\partial_\tau |\tau^\prime \rangle
= \frac{\partial}{\partial\tau}\delta(\tau-\tau^\prime)\ .
\label{partial}
\ee
We than get for the strangeness charge from the real part of
the fermion determinant in Euclidean space
\be
&&T S_R =\int\frac{d\omega}{2\pi} (i\omega)
\int_{1/\Lambda^2}^\infty ds \sqrt{\frac{s}{4\pi}}\sum_{\mu,\rho}
\langle\mu|\tilde h_{(1)}({\bf r},i\omega)|\rho\rangle
\langle\rho|\tilde h_{(1)}({\bf r},-i\omega)|\mu\rangle
\label{sr3} \\ &&\qquad\times
\Big\{R_0(s;i\omega,\epsilon_\mu,\epsilon_\rho)
-s\Big(\omega^2+(\epsilon_\mu + \epsilon_\rho )^2\Big)
\Big(R_1(s;i\omega,\epsilon_\mu,\epsilon_\rho)
-R_2(s;i\omega,\epsilon_\mu,\epsilon_\rho)\Big)\Big\}.
\nonumber
\ee
The Feynman parameter integrals are defined in eqn. (\ref{regfct}).
Note that the sum over the eigenstates $|\mu\rangle$ refers to
non-strange spinors only while the states $|\rho\rangle$ have
non-vanishing strangeness.

For the imaginary part we find
\be
S_I =\frac{1}{2T}\Tr \Big((\partial_\tau+h)^{-1}\hat S-
(\partial_\tau-h)^{-1}\hat S\Big)
=\frac{-1}{2T}\Tr \Big( \{\hat S,h\}
\Big\{\big[(\partial_\tau+h)(\partial_\tau-h)\big]^{-1}
\Big\}\Big).
\label{si1}
\ee
Proper time regularization is imposed by writing the inverse operator
in (\ref{si1}) as a parameter integral:
\be
S_I =\frac{1}{2T}\Tr \int_{1/\Lambda^2}^\infty ds \{\hat S,h\}
\exp \Big\{s\big[(\partial_\tau+h)(\partial_\tau-h)\big]
\Big\}
\label{si2}
\ee
which is identical to eqn (\ref{si1}) in the limit
$\Lambda\rightarrow\infty$. We may now apply the expansion rule
(\ref{expaexpo}) to the expression (\ref{si2}) resulting in
\be
TS_I &=&
\Tr \int_{1/\Lambda^2}^\infty ds
\hat S\big(h_{(0)}+h_{(2)}\big)\hat K_0(s)
\nonumber \\ &&
-\Tr \int_{1/\Lambda^2}^\infty sds \hat S h_{(0)}
\Big([\partial_\tau,h_{(2)}]+
\{h_{(0)},h_{(2)}\}+h_{(1)}^2\Big)\hat K_0(s)
\nonumber \\ &&
-\frac{1}{2}\Tr \int_{1/\Lambda^2}^\infty sds\ h_{(1)}
\int_0^1 dx \hat K_0(xs)\Big([\partial_\tau,h_{(1)}]+
\{h_{(0)},h_{(1)}\}\Big)\hat K_0((1-x)s)
\nonumber \\ &&
+Tr \int_{1/\Lambda^2}^\infty s^2ds \hat S h_{(0)}
\int_0^1 xdx  \hat K_0(xs)\Big([\partial_\tau,h_{(1)}]+
\{h_{(0)},h_{(1)}\}\Big)\hat K_0((1-x)s)
\nonumber \\ &&  \hspace{5cm}\times
\Big([\partial_\tau,h_{(1)}]+\{h_{(0)},h_{(1)}\}\Big)
\label{si3}
\ee
where we have made use of the identity $\{\hat S,h_{(1)}\}=h_{(1)}$.
Let us for the moment concentrate on the first term on the $RHS$
of eqn. (\ref{si3}):
\be
\frac{1}{T}\Tr \int_{1/\Lambda^2}^\infty ds
\hat S h_{(0)}\hat K_0(s) =
\frac{1}{2}\sum_{\rho=s}{\rm sign}(\epsilon_\rho)
{\rm erfc}\big|\frac{\epsilon_\rho}{\Lambda}\big|.
\label{si4}
\ee
This contribution vanishes for our field configuration
since the strange part of the Dirac sea is symmetric.
Nevertheless it is worthwhile to mention that if the Dirac
sea were distorted such that $n_s$ strange quarks
acquired negative energy eigenvalues $\epsilon_\rho<0$
we would obtain the strangeness charge $-n_s$ in the
limit $\Lambda\rightarrow\infty$. This is in agreement
with our phase convention since these states were automatically
occupied in the vacuum configuration.
For the evaluation of $S_I $ we again carry
out the temporal and spatial parts of the trace subsequently.
By repeating the manipulations leading to (\ref{sr3}) we
obtain
\be
&&T S_I = \int\frac{d\omega}{2\pi}
\int_{1/\Lambda^2}^\infty \frac{ds}{\sqrt{4\pi s}}
\sum_{\rho}e^{-s\epsilon_\rho^2}
\Big\{\big(1-2s\epsilon_\rho^2\big)\
\langle\rho|\tilde h_{(2)}({\bf r},i\omega,-i\omega)|\rho\rangle
\nonumber \\ &&\hspace{6cm}
-s\epsilon_\rho\ \langle\rho|\tilde h_{(1)}({\bf r},i\omega)
\tilde h_{(1)}({\bf r},-i\omega)|\rho\rangle\Big\}
\nonumber \\ &&\qquad
-\int\frac{d\omega}{2\pi}\int_{1/\Lambda^2}^\infty ds
\sqrt{\frac{s}{4\pi}}\sum_{\mu\rho}
\langle\mu|\tilde h_{(1)}({\bf r},i\omega)|\rho\rangle
\langle\rho|\tilde h_{(1)}({\bf r},-i\omega)|\mu\rangle
\nonumber \\ &&\hspace{1.5cm} \times
\Big\{\big(\epsilon_\mu+\epsilon_\rho\big)
R_0(s;i\omega,\epsilon_\mu,\epsilon_\rho)
-s\epsilon_\mu\big(\omega^2+(\epsilon_\mu+\epsilon_\rho)^2\big)
R_1(s;i\omega,\epsilon_\mu,\epsilon_\rho)\Big\}
\label{si5}
\ee
again $|\mu\rangle$ and $|\rho\rangle$ refer to non-strange
and strange states, respectively. Continuating back to
Minkowski space $\omega\rightarrow-i\omega$ the sum
$S_R+S_I$ yields the expression (\ref{strvac}) for $S$.

Next we wish to consider the contribution to the total spin
of the kaon bound state ${\bf J}_K$.
In order to extract the spectral function $d(\omega)$
defined in eqn (\ref{jk}) it is of course sufficient to
only consider one component of ${\bf J}_K$. For
convenience we choose this to be $J_{K3}$.
As shown in section 4) ${\bf J}_K$ is identical to the
expectation value of the grand spin operator ${\bf G}$
in the presence of the bound state. The classical
configuration has, by construction, vanishing grand spin
and does thus not contribute to ${\bf J}_K$. The valence
quark contribution is solely due to the induced strange
quark (\ref{psift}). Since for the strange quarks spin
and grand spin are identical we have
\be
T J_{K3}^{\rm val} = \eta_{\rm val}
\int\frac{d\omega}{2\pi}\sum_{\rho=s} M_\rho \frac{
\langle{\rm val}|\tilde h_{(1)}({\bf r},\omega)|\rho\rangle
\langle\rho|\tilde h_{(1)}({\bf r},-\omega)|{\rm val}\rangle}
{(\epsilon_{\rm val}-\omega-\epsilon_\rho )^2}.
\label{j3val} \\
\ee
Here $M_\rho$ is the spin projection of the intermediate
strange quark state. The determination of the Dirac sea
contribution to $J_{K3}$ is analogous to proceeding
calculation of the strangeness charge since all quark
spinors are eigenstates of the grand spin operators. We
then find in Minkowski space
\be
&&T J_{K3}^{\rm vac} = - \int\frac{d\omega}{2\pi}
\int_{1/\Lambda^2}^\infty \frac{ds}{\sqrt{4\pi s}}
\sum_{\mu=ns}M_\mu e^{-s\epsilon_\mu^2}
\Big\{\big(1-2s\epsilon_\mu^2\big)\
\langle\mu|\tilde h_{(2)}({\bf r},\omega,-\omega)|\mu\rangle
\nonumber \\ &&\hspace{6cm}
-s\epsilon_\mu\ \langle\mu|\tilde h_{(1)}({\bf r},\omega)
\tilde h_{(1)}({\bf r},-\omega)|\mu\rangle\Big\}
\label{j3vac} \\ &&\qquad
+\int\frac{d\omega}{2\pi} \int_{1/\Lambda^2}^\infty ds
\sqrt{\frac{s}{4\pi}} \sum_{\mu\nu} M_\nu
\langle\mu|\tilde h_{(1)}({\bf r},\omega)|\nu\rangle
\langle\nu|\tilde h_{(1)}({\bf r},-\omega)|\mu\rangle
(\epsilon_\mu + \epsilon_\nu - \omega)
\nonumber  \\* &&\hspace{.5cm}\times
\Big\{R_0(s; \omega , \epsilon_\mu , \epsilon_\nu)
-s(\omega+\epsilon_\mu + \epsilon_\nu)
\Big((\epsilon_\nu-\omega)
R_1(s;\omega ,\epsilon_\mu ,\epsilon_\nu)
+\omega R_2(s; \omega , \epsilon_\mu , \epsilon_\nu)
\Big)\Big\}.
\nonumber
\ee
In this evaluation of $J_{K3}^{\rm vac}$ we have used that
$\hat S \tilde h_{(2)}({\bf r},\omega,-\omega)$ and
$\hat S \tilde h_{(1)}({\bf r},\omega)\tilde h_{(1)}({\bf r},-\omega)$
are pure radial functions. These expressions therefore do
not contribute to the spin. The spin carried by the
fluctuations finally is the sum
\be
J_{K3} &=& J_{K3}^{\rm val}+J_{K3}^{\rm vac}
\label{jk3}
\ee
from which the spectral function $d(\omega)$ in eqn (\ref{jk})
may easily be extracted.

We conclude this appendix by noting that it is convenient to
define the Euclidean time interval $T$ into the fluctuations.
{\it I.e.} we substitute eqn. (\ref{annihil}) by
\be
a(r,\omega)=\sqrt{T}\eta_\omega(r)a_1(\omega)
\qquad {\rm and}\qquad
b(r,\omega)=\sqrt{T}\eta_\omega(r)a_2(\omega).
\label{deftime}
\ee
Then the overall factor $T$ drops out in eqns. (\ref{sr3}),
(\ref{si5}) and (\ref{jk3}).
\vskip 1.5cm

\stepcounter{chapter}
\leftline{\large\it Appendix B: Coupling of the Kaon Bound State
to the Collective Rotations}

\medskip

In this appendix we will derive the expression (\ref{cpar}) for the
parameter $c$ which determines the coupling between the kaon
fluctuations and the collective degrees of freedom. When expanding the
fermion determinant in terms of the angular velocity
${\mbox{\boldmath $\Omega$}}$ this coupling is obtained from the
linear expression ${\cal A}_F^{(1)}$ in
\be
{\cal A_F}={\cal A_F}({\mbox{\boldmath $\Omega$}}=0)
+\Omega_a\frac{\partial{\cal A_F}}{\partial\Omega_a}
\Big|_{{\mbox{\boldmath $\Omega$}}=0}
+{\cal O}({\mbox{\boldmath $\Omega$}}^2)
={\cal A}_F^{(0)}+{\cal A}_F^{(1)}+
{\cal O}({\mbox{\boldmath $\Omega$}}^2).
\label{expomega}
\ee
The contributions of order ${\mbox{\boldmath $\Omega$}}^2$ yield
the moment of inertia \cite{re89} which will not be discussed here.
Again we explore real and imaginary parts of the fermion
determinant separately. This calculation is most conveniently
performed in the flavor rotating system ({\it cf.} section 4).
Since we assume ${\mbox{\boldmath $\Omega$}}$ to be time
independent\footnote{In the isospin limit the $SU(2)$
rotations represent exact zero modes.} we get
\be
D^{\prime\dagger}_E D^\prime_E
=-\left(\partial_\tau^2+[\partial_\tau,h]-h^2
-i{\mbox{\boldmath $\tau$}}\cdot
{\mbox{\boldmath $\Omega$}}_E\ \partial_\tau
+\frac{i}{2}[h,{\mbox{\boldmath $\tau$}}\cdot
{\mbox{\boldmath $\Omega$}}_E]\right).
\label{expdd}
\ee
${\mbox{\boldmath $\Omega$}}_E=i{\mbox{\boldmath $\Omega$}}$
denotes the analytic continuation of angular velocities into
Euclidean space. We then obtain for the linear term in proper
time regularization
\be
{\cal A}_R^{(1)}\rightarrow
\frac{i}{2}\Omega_E^a
\Tr \int_{1/\Lambda^2}^\infty ds
\left(\tau_a\partial_\tau
+\frac{1}{2}[\tau_a,h]\right)
\exp\left(s(\partial_\tau^2-h^2+[\partial_\tau,h])\right)
\label{a1r1}
\ee
where we again made use of the fact that ${\mbox{\boldmath $\Omega$}}$
is time independent. ${\cal A}_R^{(1)}$ is now expanded up to second
order in the fluctuations according to eqn. (\ref{de2}). Denoting the
corresponding expression by ${\cal A}_R^{(1,2)}$ we find
\be
&&\hspace{-2cm}{\cal A}_R^{(1,2)}=
\frac{i}{4}\Omega_E^a \Tr \int_{1/\Lambda^2}^\infty ds
\Bigg\{\hat K_0(s)[\tau_a,h_{(2)}]
\nonumber \\&&\hspace{-1cm}
+s[\tau_a,h_{(1)}]\int_0^1 dx \hat K_0(sx)
\left([\partial_\tau,h_{(1)}]-\{h_{(0)},h_{(1)}\}\right)
\hat K_0((1-x)s)
\nonumber \\&&\hspace{-1cm}
-s\left(2\tau_a\partial_\tau+[\tau_a,h_{(0)}]\right)
\int_0^1 dx \hat K_0(sx)
\left(\{h_{(0)},h_{(2)}\}+h_{(1)}^2\right)\hat K_0((1-x)s)
\nonumber \\&&\hspace{-1cm}
+s^2\left(2\tau_a\partial_\tau+[\tau_a,h_{(0)}]\right)
\int_0^1 dx \int_0^{1-x}dy \hat K_0(sx)
\left([\partial_\tau,h_{(1)}]-\{h_{(0)},h_{(1)}\}\right)
\nonumber \\&&\hspace{1cm}\times
\hat K_0(s(1-x-y))
\left([\partial_\tau,h_{(1)}]-\{h_{(0)},h_{(1)}\}\right)
\hat K_0(sy)\Bigg\}
\label{ar12}
\ee
where we have already omitted terms which correspond to total
time derivatives. As an example let us consider the most
complicated term in more detail. Introducing eigenstates of
the Euclidean time and using eqns. (\ref{zerokernel}) and
(\ref{partial}) yields for the last term in (\ref{ar12})
\be
&&\frac{i}{4}\Omega_E^a {\it Tr} \int_{1/\Lambda^2}^\infty
\frac{ds}{4\pi}\sqrt{\frac{s}{4\pi}}
\int d\tau_1 d\tau_2 d\tau_3
\int \frac{d\omega}{2\pi}
\int \frac{d\omega^\prime}{2\pi}
\int_0^1 dx \int_0^{1-x}\frac{dy}{\sqrt{xy(1-x-y)}}
\nonumber \\ && \qquad\times
e^{-i\omega\tau_2}e^{-i\omega^\prime\tau_3}
\exp\left(-\frac{(\tau_1-\tau_2)^2}{4xs}\right)
\exp\left(-\frac{(\tau_2-\tau_3)^2}{4(1-x-y)s}\right)
\exp\left(-\frac{(\tau_1-\tau_3)^2}{4ys}\right)
\nonumber \\ && \qquad\times
\left(\tau_a\frac{\tau_1-\tau_3}{ys}+[\tau_a,h_{(0)}]\right)
e^{-sxh_{(0)}^2}
\left(-i\omega\tilde h_{(1)}(i\omega)
-\{h_{(0)}\tilde h_{(1)}(i\omega)\}\right)
\nonumber \\ &&\hspace{2cm}\times
e^{-s(1-x-y)h_{(0)}^2}
\left(-i\omega^\prime\tilde h_{(1)}(i\omega^\prime)
-\{h_{(0)}\tilde h_{(1)}(i\omega^\prime)\}\right)
e^{-syh_{(0)}^2}.
\label{ar12last}
\ee
One of the integrals $\int d\tau_i$ can be carried out trivially by
shifting $\tau_2=\tau+\tau_1$ and $\tau_3=\tau^\prime+\tau_1$ yielding
energy conservation for the fluctuations
\be
&&\hspace{-0.5cm}
\frac{i}{4}\Omega_E^a {\it Tr} \int_{1/\Lambda^2}^\infty
\frac{ds}{4\pi}\sqrt{\frac{s}{4\pi}}
\int d\tau d\tau^\prime
\int \frac{d\omega}{2\pi}
\int_0^1 dx \int_0^{1-x}\frac{dy}{\sqrt{xy(1-x-y)}}
e^{-i\omega(\tau-\tau^\prime)}
\nonumber \\ && \times
\exp\left(\frac{-1}{4s}
\left[\frac{\tau^2}{x}+\frac{\tau^{\prime2}}{y}
+\frac{(\tau-\tau^\prime)^2}{1-x-y}\right]\right)
\left(\tau_a\frac{\tau^\prime}{ys}+[\tau_a,h_{(0)}]\right)
e^{-sxh_{(0)}^2}
\label{ar12last1} \\ && \times
\left(-i\omega\tilde h_{(1)}(i\omega)
-\{h_{(0)}\tilde h_{(1)}(i\omega)\}\right)
e^{-s(1-x-y)h_{(0)}^2}
\left(i\omega\tilde h_{(1)}(-i\omega)
-\{h_{(0)}\tilde h_{(1)}(-i\omega)\}\right)
e^{-syh_{(0)}^2}.
\nonumber
\ee
The Gaussian integrals are evaluated by shifting
$\tau=\tilde\tau+\frac{x\tau^\prime}{1-y}$ and integrating
over $\tilde\tau$. Subsequently we introduce
\footnote{There are no poles in the $\tau^\prime$ plane.}
$\tau^\prime=\tilde\tau^\prime+2isy(1-x-y)\omega$
and perform the remaining integral $\int d\tilde\tau^\prime$.
This results in
\be
&&\frac{i}{4}\Omega_E^a \int_{1/\Lambda^2}^\infty
ds\sqrt{\frac{s^3}{4\pi}} \int \frac{d\omega}{2\pi}
\int_0^1 dx \int_0^{1-x} dy
\sum_{\mu,\nu=n.s.  \atop \rho=s}
\left[2i\omega(1-x-y)+\epsilon_\nu-\epsilon_\mu\right]
\left(\epsilon_\nu+\epsilon_\rho+i\omega\right)
\nonumber \\ && \quad \times
\left(\epsilon_\mu+\epsilon_\rho-i\omega\right)
\exp\left(-s\left[(1-x-y)(x+y)\omega^2+(1-x-y)\epsilon_\rho^2
+x\epsilon_\nu^2+y\epsilon_\mu^2\right]\right)
\nonumber \\ && \hspace{3cm}\times
\langle\mu|\tau_a|\nu\rangle
\langle\nu|\tilde h_{(1)}(i\omega)|\rho\rangle
\langle\rho|\tilde h_{(1)}(-i\omega)|\mu\rangle
\label{ar12last2}
\ee
where we have evaluated the trace over the spatial and internal
degrees of freedom in the eigenbasis of $h_{(0)}$ (\ref{diagham}).
We have furthermore used the fact that $\tau_a$ have non-vanishing
matrix elements between non-strange states only. Redefining the
Feynman parameters $\alpha=x+y$ and $\beta=x-y$ we see that
$\beta$ appears only linearly in the exponent. Thus this integral
may be carried out analytically
\be
&&\frac{-i}{4}\Omega_E^a \int_{1/\Lambda^2}^\infty
ds\sqrt{\frac{s}{4\pi}} \int \frac{d\omega}{2\pi}
\int_0^1 d\alpha \sum_{\mu,\nu=n.s.  \atop \rho=s}
\left[2i\omega(1-\alpha)+\epsilon_\nu-\epsilon_\mu\right]
\left(\epsilon_\nu+\epsilon_\rho+i\omega\right)
\nonumber \\&& \qquad \times
\left(\epsilon_\mu+\epsilon_\rho-i\omega\right)
\exp\left(-s(1-\alpha)(\alpha\omega^2+\epsilon_\rho^2)\right)
\frac{e^{-s\alpha\epsilon_\mu^2}-e^{-s\alpha\epsilon_\nu^2}}
{\epsilon_\mu^2-\epsilon_\nu^2}
\nonumber \\&& \hspace{3cm} \times
\langle\mu|\tau_a|\nu\rangle
\langle\nu|\tilde h_{(1)}(i\omega)|\rho\rangle
\langle\rho|\tilde h_{(1)}(-i\omega)|\mu\rangle.
\label{ar12last3}
\ee
The remaining terms of ${\cal A}_R^{(1,2)}$ in eqn. (\ref{ar12})
are treated similarly.

Next we turn to the contribution stemming from the imaginary
part. For the expression linear in the angular velocity we
get
\be
{\cal A}_I^{(1)}=\frac{1}{2}\Omega_a \Tr
\frac{\partial}{\partial\Omega_a}
\left({\rm log}D_E^\prime-{\rm log}
D_E^{\prime\dagger}\right)
\Big|_{{\mbox{\boldmath $\Omega$}}=0}
=\frac{i}{4}\Omega_a \Tr
\left\{\left[(\partial_\tau-h)(\partial_\tau+h)\right]^{-1}
\{h,\tau_a\}\right\}.
\label{ai1}
\ee
As for the calculation of the strangeness charge the proper time
cut-off $\Lambda$ is introduced by writing the inverse operator
in (\ref{ai1}) as a parameter integral
\be
{\cal A}_I^{(1)}\rightarrow\frac{-i}{4}\Omega_a \Tr
\{h,\tau_a\}\int_{1/\Lambda^2}^\infty ds
\exp\left(s(\partial_\tau-h)(\partial_\tau+h)\right)
\label{ai1cut}
\ee
To this expression we then apply the general formula (\ref{expaexpo})
and obtain the analogue to eqn. (\ref{ar12})
\be
&&\hspace{-2cm}
{\cal A}_I^{(1,2)}=
-\frac{i}{4}\Omega_E^a \Tr \int_{1/\Lambda^2}^\infty ds
\Bigg\{\hat K_0(s)\{h_{(2)},\tau_a\}
\nonumber \\ &&\qquad
-s\{h_{(0)},\tau_a\}\int_0^1 dx \hat K_0(xs)
\left(\{h_{(0)},h_{(2)}\}+h_{(1)}^2\right)\hat K_0((1-x)s)
\nonumber \\ &&\qquad
+s\{h_{(1)},\tau_a\}\int_0^1 dx \hat K_0(xs)
\left([\partial_\tau,h_{(1)}]-\{h_{(0)},h_{(1)}\}\right)
\hat K_0((1-x)s)
\nonumber \\ &&\qquad
+s^2\{h_{(0)},\tau_a\}\int_0^1 dx\int_0^{1-x} dy \hat K_0(xs)
\left([\partial_\tau,h_{(1)}]-\{h_{(0)},h_{(1)}\}\right)
\nonumber \\ && \hspace{3cm}\times
\hat K_0((1-x-y)s)
\left([\partial_\tau,h_{(1)}]-\{h_{(0)},h_{(1)}\}\right)
\hat K_0(ys)\Bigg\}.
\label{ai12}
\ee
We finally get for the sum
${\cal A}_F^{(1,2)}={\cal A}_R^{(1,2)}+{\cal A}_I^{(1,2)}$
\be
&&\hspace{-1.5cm}
{\cal A}_F^{(1,2)}=\frac{i}{2}\Omega_E^a
\int_{1/\Lambda^2}^\infty \frac{ds}{\sqrt{4\pi s}}
\int \frac{d\omega}{2\pi}\Bigg\{
\sum_{\mu,\nu=n.s.}\langle\mu|\tau_a|\nu\rangle
\langle\nu|\tilde h_{(2)}(i\omega,-i\omega)|\mu\rangle
\frac{\epsilon_\mu e^{-s\epsilon_\mu^2}
-\epsilon_\nu e^{-s\epsilon_\nu^2}}
{\epsilon_\mu-\epsilon_\nu}
\nonumber \\ && \hspace{-0.5cm}
+\sum_{\mu,\nu=n.s. \atop \rho=s.}
\langle\mu|\tau_a|\nu\rangle\langle\nu|
\tilde h_{(1)}(i\omega)|\rho\rangle
\langle\rho|\tilde h_{(1)}(-i\omega)|\mu\rangle
\Big[\epsilon_\nu
\frac{e^{-s\epsilon_\mu^2}-e^{-s\epsilon_\nu^2}}
{\epsilon_\mu^2-\epsilon_\nu^2}+
s{\cal R}_{\mu,\nu,\rho}(i\omega)\Big]\Bigg\}.
\label{af12}
\ee
Taking advantage of the symmetry under
$\epsilon_\mu\leftrightarrow\epsilon_\nu$ yields
the regulator function
\be
&&\hspace{-1.5cm}
{\cal R}_{\mu,\nu,\rho}(\omega)=
\int_0^1 dx \Bigg\{(\omega+\epsilon_\nu+\epsilon_\rho)
\exp\left(-s[(1-x)\epsilon_\nu^2
+x\epsilon_\rho^2-x(1-x)\omega^2]\right)
\nonumber \\ && \hspace{-0.5cm}
+\Big(2(1-x)\omega^3+[\epsilon_\mu+\epsilon_\nu]\omega^2
-[2(1-x)(\epsilon_\nu+\epsilon_\rho)(\epsilon_\mu+\epsilon_\rho)
-(\epsilon_\mu-\epsilon_\nu)^2]\omega
\label{af12r} \\ && \hspace{-0.5cm}
+(\epsilon_\mu+\epsilon_\nu)
(\epsilon_\mu+\epsilon_\rho)(\epsilon_\nu+\epsilon_\rho)\Big)
\frac{e^{-s\epsilon_\mu^2}-e^{-s\epsilon_\nu^2}}
{\epsilon_\mu^2-\epsilon_\nu^2}
\exp\left(-s[(1-x)\epsilon_\rho^2-x(1-x)\omega^2]\right)
\Bigg\}. \nonumber
\ee

The valence quark contribution is finally obtained by investigating
the expression
\be
{\cal A}_{\rm val}=\int d^4x
\left(\Psi_{\rm val}^{ns}({\bf r},t)
+\delta \Psi_{\rm val}({\bf r},t)\right)^\dagger
\left[i\partial_t-h-\frac{1}{2}
{\mbox{\boldmath $\tau$}}\cdot{\mbox{\boldmath $\Omega$}}
\right]
\left(\Psi_{\rm val}^{ns}({\bf r},t)
+\delta \Psi_{\rm val}({\bf r},t)\right).
\label{a12val}
\ee
Since the valence quark part of the action does not undergo
regularization there is no need to continue back and forth
between Euclidean and Minkowski space, {\it i.e.} all
calculations involving ${\cal A}_{\rm val}$ are carried out
in Minkowski space. The perturbation of the valence quark
wave-function contains both strange and non-strange parts
\be
\delta \Psi_{\rm val}({\bf r},t)=
\delta \Psi_{\rm val}^s({\bf r},t)+
\delta \Psi_{\rm val}^{ns}({\bf r},t).
\label{delpsi}
\ee
$\delta \Psi_{\rm val}^s({\bf r},t)$ is given in eqn. (\ref{psistr})
while  the non-strange piece is due to the collective rotation
\be
\delta \Psi_{\rm val}^{ns}({\bf r},t)=\frac{1}{2}
\sum_{\mu\ne{\rm val}}\psi_\mu({\bf r})
\frac{\langle \mu |{\mbox{\boldmath $\tau$}}
\cdot{\mbox{\boldmath $\Omega$}}|{\rm val}\rangle}
{\epsilon_{\rm val}-\epsilon_\mu}
e^{-i\epsilon_{\rm val}t}.
\label{delpsins}
\ee
${\cal A}_{\rm val}$ is expanded up to quadratic order in the
fluctuations and linear in the angular velocities:
\be
{\cal A}_{\rm val}^{(1,2)}=-\int d^4x\Big\{
\psi_{\rm val}^\dagger h_{(2)}\delta \Psi_{\rm val}^{ns}
+(\delta \Psi_{\rm val}^{ns})^\dagger h_{(1)}
\delta \Psi_{\rm val}^s \ + \ {\rm h.\ c.}\Big\}
\label{aval12}
\ee
since $(\delta \Psi_{\rm val}^s)^\dagger
{\mbox{\boldmath $\tau$}}\cdot{\mbox{\boldmath $\Omega$}}
\delta \Psi_{\rm val}^s=0$. Substitution of eqns.
(\ref{psistr},\ref{delpsins}) leads to
\be
&&\hspace{-1.5cm}
{\cal A}_{\rm val}^{(1,2)}=-\frac{1}{2}\int \frac{d\omega}{2\pi}
\Bigg\{\sum_{\mu=ns}\Big[\langle{\rm val}|
\tilde h_{(2)}(\omega,-\omega)|\mu\rangle\langle\mu|
{\mbox{\boldmath $\tau$}}\cdot{\mbox{\boldmath $\Omega$}}
|{\rm val}\rangle \frac{1}{\epsilon_{\rm val}-\epsilon_\mu}
\ + \ {\rm h.\ c.}\Big]
\label{aval12a} \\ && \hspace{-1.0cm}
+\sum_{\mu=ns \atop \rho=s}\Big[
\langle{\rm val}|\tilde h_{(1)}(\omega)|\rho\rangle
\langle\rho|\tilde h_{(1)}(-\omega)|\mu\rangle\langle\mu|
{\mbox{\boldmath $\tau$}}\cdot{\mbox{\boldmath $\Omega$}}
|{\rm val}\rangle \frac{1}{(\epsilon_{\rm val}-\epsilon_\mu)
(\epsilon_{\rm val}-\omega-\epsilon_\rho)}
+ \ {\rm h.\ c.}\Big]\Bigg\}.
\nonumber
\ee
This result together with that for the Dirac sea contribution
(\ref{af12}) continued back to Minkowski space allows to read off
the parameter $c$ introduced in section 4. According to the
definition (\ref{deftime}) ${\cal A}^{(1,2)}$ acquires an
overall factor $T$ which provides the possibility to read off
those parts of the collective Lagrangian which correspond
to ${\cal A}^{(1,2)}$.
\vskip 1.5cm

\stepcounter{chapter}
\leftline{\large\it Appendix C: Matrix Elements for the Kaon Bound
State}

\medskip

In this appendix we will present the actual evaluation of the
matrix elements $\langle \mu | \tilde h_i(\omega) |\nu \rangle$
appearing in the expansion of the fermion determinant.

\smallskip
\leftline{\it C.1 General definitions}

We start by listing the basis spinors used to diagonalize the
zeroth-order Hamiltonian $h_{(0)}$ (\ref{h0}). $h_{(0)}$ commutes with
the grand spin operator
\be
{\mbox{\boldmath $G$}}=
{\mbox{\boldmath $J$}}+\frac{{\mbox{\boldmath $\tau$}}}{2}
={\mbox{\boldmath $l$}}+\frac{{\mbox{\boldmath $\sigma$}}}{2}
+\frac{{\mbox{\boldmath $\tau$}}}{2}
\label{gspin}
\ee
which is the sum of the total spin ${\mbox{\boldmath $J$}}$ and the
isospin ${\mbox{\boldmath $\tau$}}/2$. The spin itself is
decomposed into orbital angular momentum ${\mbox{\boldmath $l$}}$
and intrinsic spin ${\mbox{\boldmath $\sigma$}}/2$.
It is then useful to construct eigenstates of
${\mbox{\boldmath $G$}}$ with eigenvalues $G$ and the
corresponding projection quantum number $M$:
\be
|l j G M\rangle,\quad j=\cases{G+1/2, & $l=\cases{G+1 &\cr G &}$ \cr
& \cr G-1/2, & $l=\cases{G &\cr G-1 &}$}.
\label{gstates}
\ee
The eigenvalues of ${\mbox{\boldmath $J$}}$ and
${\mbox{\boldmath $l$}}$ are denoted by $j$ and $l$, respectively.
Now we easily find the grand spin symmetric eigenfunctions of the
free Hamiltonian ({\it i.e.} $\Theta=0$ in eqn.(\ref{h0})) in the
non-strange sector:
\be
|1,n,j=G+\frac{1}{2},M\rangle &=& {\cal N}_n^G
\pmatrix{iw^+_{nG}j_G(k_{nG}r)|GG+\frac{1}{2}GM\rangle \cr
w^-_{nG}j_{G+1}(k_{nG}r)|G+1G+\frac{1}{2}GM\rangle \cr},\nonumber \\*
e^0&=&E_{nG}=\pm\sqrt{k_{nG}^2+m^2},\quad \Pi=(-)^G.\nonumber \\*
\nonumber \\*
|2,n,j=G-\frac{1}{2},M\rangle &=& {\cal N}_n^G
\pmatrix{iw^+_{nG}j_G(k_{nG}r)|GG-\frac{1}{2}GM\rangle \cr
-w^-_{nG}j_{G-1}(k_{nG}r)|G-1G-\frac{1}{2}GM\rangle \cr},\nonumber \\*
e^0&=&E_{nG}=\pm\sqrt{k_{nG}^2+m^2},\quad \Pi=(-)^G.\nonumber \\*
\nonumber \\*
|3,n,j=G+\frac{1}{2},M\rangle &=& {\cal N}_n^{G+1}
\pmatrix{iw^+_{nG+1}j_{G+1}(k_{nG+1}r)|G+1G+\frac{1}{2}GM\rangle \cr
-w^-_{nG+1}j_G(k_{nG+1}r)|GG+\frac{1}{2}GM\rangle \cr},\nonumber \\*
e^0&=&E_{nG+1}=\pm\sqrt{k_{nG+1}^2+m^2},\quad \Pi=(-)^{G+1}.\nonumber
\\*
\nonumber \\*
|4,n,j=G-\frac{1}{2},M\rangle &=& {\cal N}_n^{G-1}
\pmatrix{iw^+_{nG-1}j_{G-1}(k_{nG-1}r)|G-1G-\frac{1}{2}GM\rangle \cr
w^-_{nG-1}j_G(k_{nG-1}r)|GG-\frac{1}{2}GM\rangle \cr},\nonumber \\*
e^0&=&E_{nG-1}=\pm\sqrt{k_{nG-1}^2+m^2},\quad \Pi=(-)^{G+1}.
\label{basis}
\ee
wherein $j_l$ denote the spherical Bessel functions. Also the
energy eigenvalues $e^0$ with respect to the free Hamiltonian and
the parity quantum numbers $\Pi$ are listed. The kinematical
factors $w_{nl}^\pm$ are defined as
\be
w^+_{nl}&=&\sqrt{1+m/E_{nl}},\quad
w^-_{nl}={\rm sign}(E_{nl})\sqrt{1-m/E_{nl}}.
\label{kine}
\ee
This system is treated in a spherical box of finite radius $D$.
Thereby the momentum eigenvalues are
discretized by demanding the upper
components to vanish at the boundary, {\it i.e.}
\be
j_l(k_{nl}D)=0.
\label{quantk}
\ee
These boundary conditions have been noticed to be suitable for
the evaluation of matrix elements of flavor generators as {\it e.g.}
$\tau_3$\cite{we92}. Then the normalization is given by
\be
{\cal N}_n^G=\big[D^{3/2}|j_{G+1}(k_{nG}D)|\big]^{-1}.
\label{norm}
\ee

The numerical calculations require
an upper bound for the momenta $k_{nl}$ \footnote{This upper
bound is typically chosen four times as large as the
cut-off $\Lambda$.} leaving for each $l$ a specified number
of momentum eigenstates $N(l)$: $k_{1l},...,k_{N(l)l}$.
Taking into account that for each momentum one has two
energy eigenvalues $e^0$ it is obvious that we have $2N(l)$
eigenfunctions of $h_0(\Theta=0)$ for each grand spinor in
eqn. (\ref{basis}).

We can now easily diagonalize the zeroth-order Hamiltonian $h_0$
in the isospace subgroup. The corresponding eigenstates are
linear combinations of the spinors in (\ref{basis}). The
eigenstates with parity $\Pi=(-)^{G}$ read:
\be
|\mu, G, M, + \rangle = &&
\sum_{n=1}^{2N(G)}V_G^{(+)}(n,\mu)|1,n,G,M\rangle
\nonumber \\
&&+\sum_{n=1}^{2N(G)}V_G^{(+)}(n+2N(G),\mu)|2,n,G,M\rangle
\label{poseig}
\ee
while those with parity $\Pi=(-)^{G+1}$ are given by
\be
|\mu, G, M, - \rangle =&&
\sum_{n=1}^{2N(G+1)}V_G^{(-)}(n,\mu)|3,n,G,M\rangle
\nonumber \\
&&+\sum_{n=1}^{2N(G-1)}V_G^{(-)}(n+2N(G+1),\mu)|4,n,G,M\rangle.
\label{negeig}
\ee
In general $V_G^{(+)}$ is a $4N(G)\times4N(G)$ orthogonal matrix
while $V_G^{(-)}$ is $2(N(G+1)+N(G-1))\times2(N(G+1)+N(G-1))$
dimensional. For $G=0$, however, the states $|2,n,G,M\rangle$
and $|4,n,G,M\rangle$ do not exist and $V_0^{(\pm)}$ is
reduced accordingly.

The matrix elements $\langle\mu|\tau_3|\nu\rangle$
are straightforwardly evaluated since the grand spin eigenstates
are obtained by coupling states with good spin $|ljm\rangle$
and isospin $|1/2 m_t\rangle$
\be
|ljGM\rangle = \sum_mC^{GM}_{jm,1/2M-m}|ljm\rangle|1/2M-m\rangle.
\label{Coup1}
\ee
Substituting the relevant Clebsch-Gordan coefficients yields
\be
&&\langle\mu,G,M,+|\tau_3|\nu,G^\prime,M^\prime,+\rangle=
\nonumber \\ && \quad
-\delta_{GG^\prime}\delta_{MM^\prime}
\sum_{n=1}^{2N(G)}\Big\{
\frac{M}{G+1}V_G^{(+)}(n,\mu)V_G^{(+)}(n,\nu)
\nonumber \\ && \qquad
-\frac{M}{G}V_G^{(+)}(n+2N(G),\mu)
V_G^{(+)}(n+2N(G),\nu)\Big\},
\label{tau3pp} \\
\vspace{0.5cm}
&&\langle\mu,G,M,+|\tau_3|\nu,G^\prime,M^\prime,-\rangle=
\nonumber \\ && \quad
-\delta_{MM^\prime}\sum_{n=1}^{2N(G)}\Big\{
\delta_{G+1G^\prime}\sqrt{1-\left(\frac{M}{G+1}\right)^2}
V_G^{(+)}(n,\mu)V_{G+1}^{(-)}(n+2N(G+2),\nu)
\nonumber \\ && \qquad
+\delta_{G-1G^\prime}\sqrt{1-\left(\frac{M}{G}\right)^2}
V_G^{(+)}(n+2N(G+2),\mu)V_{G-1}^{(-)}(n+2N(G+2),\nu)\Big\},
\label{tau3pm} \\
\vspace{0.5cm}
&&\langle\mu,G,M,-|\tau_3|\nu,G^\prime,M^\prime,+\rangle=
\nonumber \\ && \quad
-\delta_{MM^\prime}\sum_{n=1}^{2N(G^\prime)}\Big\{
\delta_{GG^\prime-1}\sqrt{1-\left(\frac{M}{G^\prime}\right)^2}
V_{G^\prime-1}^{(-)}(n,\mu)V_{G^\prime}^{(+)}(n+2N(G^\prime),\nu)
\nonumber \\ && \qquad
+\delta_{GG^\prime+1}\sqrt{1-\left(\frac{M}{G^\prime+1}\right)^2}
V_{G^\prime+1}^{(-)}(n+2N(G^\prime+2),\mu)
V_{G^\prime}^{(+)}(n,\nu)\Big\},
\label{tau3mp} \\
\vspace{0.5cm}
&&\langle\mu,G,M,-|\tau_3|\nu,G^\prime,M^\prime,-\rangle=
\nonumber \\ && \quad
-\delta_{GG^\prime}\delta_{MM^\prime}
\Big\{ \frac{M}{G+1} \sum_{n=1}^{2N(G+1)}
V_G^{(-)}(n,\mu)V_G^{(-)}(n,\nu)
\nonumber \\ && \qquad
-\frac{M}{G}\sum_{n=1}^{2N(G+1)}V_G^{(-)}(n+2N(G+1),\mu)
V_G^{(-)}(n+2N(G+1),\nu)\Big\}.
\label{tau3mm}
\ee

For the evaluation of the matrix elements including the
kaon bound state it is convenient to define radial
functions which correspond to the radial part ot the
eigenstates:
\be
g_\mu^{(G,+;1)}(r)&=&\sum_{n=1}^{2N(G)}V_G^{(+)}(n,\mu){\cal N}_n^G
w^+_{nG}j_G(k_{nG}r)
\nonumber \\
g_\mu^{(G,+;2)}(r)&=&\sum_{n=1}^{2N(G)}V_G^{(+)}(n+2N(G),\mu)
{\cal N}_n^G w^+_{nG}j_G(k_{nG}r)
\nonumber \\
f_\mu^{(G,+;1)}(r)&=&\sum_{n=1}^{2N(G)}V_G^{(+)}(n,\mu){\cal N}_n^G
w^-_{nG}j_{G+1}(k_{nG}r)
\nonumber \\
f_\mu^{(G,+;2)}(r)&=&\sum_{n=1}^{2N(G)}V_G^{(+)}(n+2N(G),\mu)
{\cal N}_n^G w^-_{nG}j_{G-1}(k_{nG}r)
\label{efctpos}
\ee
Then the eigenstates of $h_0$ with grand spin $G$ and parity
$\Pi=(-)^G$ are given by
\be
\Psi_\mu^{(G,+)}=
\pmatrix{ig_\mu^{(G,+;1)}(r)|GG+\frac{1}{2}GM\rangle \cr
f_\mu^{(G,+;1)}(r) |G+1G+\frac{1}{2}GM\rangle \cr} +
\pmatrix{ig_\mu^{(G,+;2)}(r)|GG-\frac{1}{2}GM\rangle \cr
-f_\mu^{(G,+;2)}(r) |G-1G-\frac{1}{2}GM\rangle \cr}
\label{psipos}
\ee
Analogously to eqn. (\ref{efctpos}) radial functions
$g_\mu^{(G,-;1)}(r),g_\mu^{(G,-;2)}(r),f_\mu^{(G,-;1)}(r)$
and $f_\mu^{(G,-;2)}(r)$ are defined such that the
eigenstates of $h_0$ with grand spin $G$ and parity
$\Pi=(-)^{G+1}$ are given by
\be
\Psi_\mu^{(G,-)}=
\pmatrix{ig_\mu^{(G,-;1)}(r)|G+1G+\frac{1}{2}GM\rangle \cr
-f_\mu^{(G,-;1)}(r) |GG+\frac{1}{2}GM\rangle \cr} +
\pmatrix{ig_\mu^{(G,-;2)}(r)|G-1G-\frac{1}{2}GM\rangle \cr
f_\mu^{(G,-;2)}(r) |GG-\frac{1}{2}GM\rangle \cr}
\label{psineg}
\ee

The strange quarks have vanishing isospin. Thus their grand
spin is identical to their ordinary spin. We classify the
basis spinors by their total spin $j$ and orbital angular
momentum $l$:
\be
|1,\rho,j=l+\frac{1}{2},m\rangle _{\rm s}&=&{\cal N}_\rho^l
\pmatrix{i{\overline{w}}_{\rho l}^{\hskip 0.3em+}
j_l(k_{\rho l}r)|ljm\rangle \cr
{\overline{w}}_{\rho l}^{\hskip 0.3em-}
j_{l+1}(k_{\rho l}r)|l+1jm\rangle
\cr},\nonumber
\\*
e^0&=&{\overline{E}}_{\rho l}=\pm\sqrt{k_{\rho l}^2+m_s^2}.
\nonumber \\*
|2,\rho,j=l-\frac{1}{2},m\rangle _{\rm s}&=&{\cal N}_\rho^l
\pmatrix{i{\overline{w}}_{\rho l}^{\hskip 0.3em+}
j_l(k_{\rho l}r)|ljm\rangle \cr
-{\overline{w}}_{\rho l}^{\hskip 0.3em-}
j_{l-1}(k_{\rho l}r)|l-1jm\rangle \cr},
\nonumber \\*
e^0&=&{\overline{E}}_{\rho l}=\pm\sqrt{k_{\rho l}^2+m_s^2}.
\nonumber \\* \nonumber \\*
{\overline{w}}_{\rho l}^{\hskip 0.3em+}&=&\sqrt{1+m_s/\hskip 0.3em
{\overline{E}}_{\rho l}},\quad {\overline{w}}_{\rho l}^{\hskip 0.3em-}=
{\rm sign}({\overline{E}}_{\rho l})
\sqrt{1-m_s/\hskip 0.3em {\overline{E}}_{\rho l}}.
\label{stbasis}
\ee
Note the appearance of the strange constituent mass $m_s$ in the
energy eigenvalues. Since the chiral soliton has vanishing
strangeness the eigenstates (\ref{stbasis}) of the free
Hamiltonian are also eigenstates of the zeroth-order Hamiltonian
(\ref{h0}).

According to the discussion in section 3 we may write the
perturbative parts $h_{(i)}$, $i=1,2$ which contain the kaon
bound state wave-function in the form (in flavor space):
\be
\tilde h_{(1)}(\omega)=-\frac{1}{2}(m+m_s)\pmatrix{u_0 & 0 \cr 0 &1\cr}
\pmatrix{0 & \Omega(\omega,r)\cr \Omega^\dagger(-\omega,r) &0\cr}
\pmatrix{u_0 & 0 \cr 0 &1\cr}
\label{h1u0}
\ee
$\Omega(\omega,r)$ denotes the two-component isospinor
defined in eqn.(\ref{bound}) and the unitary, self-adjoint
transformation matrix $u_0$ is defined in eqn. (\ref{u0}).
In the same way $h_{(2)}$ may be written as
\be
\tilde h_{(2)}(\omega)=\frac{1}{4}(m+m_s)
\pmatrix{u_0 & 0 \cr 0 &1\cr}\beta
\pmatrix{\Omega(\omega,r)\Omega^\dagger(\omega,r) &0 \cr
0& \hspace{-1cm}\Omega^\dagger(-\omega,r)\Omega(-\omega,r) \cr}
\pmatrix{u_0 & 0 \cr 0 &1\cr}.
\label{h2u0}
\ee
Acting with $u_0$ on $\Psi_\mu^{G\pm}$ defines new radial
functions $\tilde g_\mu^{G,+;1}(r)$, etc. via
\be
u_0 \Psi_\mu^{(G,+)}=
\pmatrix{i\tilde g_\mu^{(G,+;1)}(r)|GG+\frac{1}{2}GM\rangle \cr
\tilde f_\mu^{(G,+;1)}(r) |G+1G+\frac{1}{2}GM\rangle \cr} +
\pmatrix{i\tilde g_\mu^{(G,+;2)}(r)|GG-\frac{1}{2}GM\rangle \cr
-\tilde f_\mu^{(G,+;2)}(r) |G-1G-\frac{1}{2}GM\rangle \cr},
\nonumber \\
\nonumber \\
u_0 \Psi_\mu^{(G,-)}=
\pmatrix{i\tilde g_\mu^{(G,-;1)}(r)|G+1G+\frac{1}{2}GM\rangle \cr
-\tilde f_\mu^{(G,-;1)}(r) |GG+\frac{1}{2}GM\rangle \cr} +
\pmatrix{i\tilde g_\mu^{(G,-;2)}(r)|G-1G-\frac{1}{2}GM\rangle \cr
\tilde f_\mu^{(G,-;2)}(r) |GG-\frac{1}{2}GM\rangle \cr}.
\label{psirot}
\ee
As an example we display
\be
\tilde g_\mu^{(G,+;1)}(r)=
{\rm sin}\frac{\Theta}{2}g_\mu^{(G,+;1)}(r)
-\frac{1}{2G+1}{\rm cos}\frac{\Theta}{2}f_\mu^{(G,+;1)}(r)
-\frac{2\sqrt{G(G+1)}}{2G+1}
{\rm cos}\frac{\Theta}{2}f_\mu^{(G,+;2)}(r).
\nonumber
\ee

\smallskip
\leftline{\it C.2 The  Matrix elements involving
$\tilde h_{(1)}(\omega)$}

Employing the radial functions $\tilde g_\mu^{(G,+;1)}(r)$ etc.
the matrix elements $\langle\mu|\tilde h_1(\omega)|\rho{\rangle_s}$
may be displayed compactly:
\be
&&\langle \mu,G,M,+|\tilde h_{(1)}(\omega)| 1,\rho,l,m{\rangle_s}=
-\frac{m+m_s}{2}\delta_{Gl}{\cal N}_\rho^G
\nonumber \\
&&\qquad \times
\int_0^D dr r^2
\Big[{\overline w}_{\rho G}^{\hskip 0.3em+}j_{G}(k_{\rho G}r)
\tilde g_\mu^{(G,+;1)}(r)+
{\overline w}_{\rho G}^{\hskip 0.3em-}j_{G+1}(k_{\rho G}r)
\tilde f_\mu^{(G,+;1)}(r)\Big]
\nonumber \\
&&\qquad \times
\Big[-\sqrt{\frac{G-M+1}{2G+2}}\delta_{Mm+1/2}a(r,\omega)
+\sqrt{\frac{G+M+1}{2G+2}}\delta_{Mm-1/2}b(r,\omega)\Big],
\label{h1p1} \\
\vspace{0.5cm}
&&\langle \mu,G,M,+|\tilde h_{(1)}(\omega)| 2,\rho,l,m{\rangle_s}=
-\frac{m+m_s}{2}\delta_{Gl}{\cal N}_\rho^{G}
\nonumber \\
&&\qquad \times
\int_0^D dr r^2
\Big[{\overline w}_{\rho G}^{\hskip 0.3em+}j_{G}(k_{\rho G}r)
\tilde g_\mu^{(G,+;2)}(r)+
{\overline w}_{\rho G}^{\hskip 0.3em-}j_{G-1}(k_{\rho G}r)
\tilde f_\mu^{(G,+;2)}(r)\Big]
\nonumber \\
&&\qquad \times
\Big[\sqrt{\frac{G+M}{2G}}\delta_{Mm+1/2}a(r,\omega)
+\sqrt{\frac{G-M}{2G}}\delta_{Mm-1/2}b(r,\omega)\Big],
\label{h1p2} \\
\vspace{0.5cm}
&&\langle \mu,G,M,-|\tilde h_{(1)}(\omega)| 1,\rho,l,m{\rangle_s}=
-\frac{m+m_s}{2}\delta_{G-1l}{\cal N}_\rho^{G-1}
\nonumber \\
&&\qquad \times
\int_0^D dr r^2
\Big[{\overline w}_{\rho G-1}^{\hskip 0.3em+}j_{G-1}(k_{\rho G-1}r)
\tilde g_\mu^{(G,-;2)}(r)+
{\overline w}_{\rho G-1}^{\hskip 0.3em-}j_{G}(k_{\rho G-1}r)
\tilde f_\mu^{(G,-;2)}(r)\Big]
\nonumber \\
&&\qquad \times
\Big[\sqrt{\frac{G+M}{2G}}\delta_{Mm+1/2}a(r,\omega)
+\sqrt{\frac{G-M}{2G}}\delta_{Mm-1/2}b(r,\omega)\Big],
\label{h1m1} \\
\vspace{0.5cm}
&&\langle \mu,G,M,-|\tilde h_{(1)}(\omega)| 2,\rho,l,m{\rangle_s}=
-\frac{m+m_s}{2}\delta_{G+1l}{\cal N}_\rho^{G+1}
\nonumber \\
&&\qquad \times
\int_0^D dr r^2
\Big[{\overline w}_{\rho G+1}^{\hskip 0.3em+}j_{G+1}(k_{\rho G+1}r)
\tilde g_\mu^{(G,-;1)}(r)+
{\overline w}_{\rho G+1}^{\hskip 0.3em-}j_{G}(k_{\rho G+1}r)
\tilde f_\mu^{(G,-;1)}(r)\Big]
\nonumber \\
&&\qquad \times
\Big[-\sqrt{\frac{G-M+1}{2G}}\delta_{Mm+1/2}a(r,\omega)
+\sqrt{\frac{G+M+1}{2G}}\delta_{Mm-1/2}b(r,\omega)\Big].
\label{h1m2}
\ee
In the representation (\ref{psirot}) the matrix elements of
$\tilde h_2(\omega,-\omega)$ reduce to matrix elements of the
form
$(a(r,\omega)a^*(r,\omega)+b(r,\omega)b^*(r,\omega))\beta$
which are trivially evaluated.

\smallskip
\leftline{\it C.3 The  Matrix elements involving
$\tilde h_{(1)}(\omega) \times \tilde h_{(1)}(-\omega)$}

The matrix elements $\langle\mu|\tilde h_{(1)}(\omega)|\rho{\rangle_s}
{_s\langle}\rho|\tilde h_{(1)}(-\omega)|\mu\rangle$ are easily
obtained using eqns. (\ref{psirot}). Making furthermore
use of eqns. (\ref{tau3pp}-\ref{tau3mm}) we are enabled to
evaluate the matrix elements $\langle \mu |\tau_3|\nu\rangle
\langle\nu|\tilde h_{(1)}(\omega)|\rho{\rangle_s}
{_s\langle}\rho|\tilde h_{(1)}(-\omega)|\mu\rangle$. Of course, the
energy eigenvalues of $h_{(0)}$ are degenerate with respect to the
grand spin projection quantum number $M$. The corresponding sum
may thus be carried out regardless of the regulator function. We define
\be
&&\sum_{M_\mu,M_\nu,M_\rho}\langle \mu |\tau_3|\nu\rangle
\langle\nu|\tilde h_{(1)}(\omega)|\rho{\rangle_s}
{_s\langle}\rho|\tilde h_{(1)}(-\omega)|\mu\rangle=
\nonumber \\
&&\qquad\qquad\qquad \frac{(m+m_s)^2}{24}
\Gamma_{\mu\nu\rho}(G_\mu,\pi_\mu;G_\nu,\pi_\nu;l_\rho,\sigma_\rho)
\label{defgamma}
\ee
$G_\mu,M_\mu$ and $\pi_\mu=\pm$ label the non-strange spinors of
eqns. (\ref{psipos},\ref{psineg}):
$|\mu\rangle =|\mu, G_\mu,M_\mu,\pi_\mu\rangle$
The strange spinors of eqn. (\ref{stbasis})
are characterized by $l_\rho$ and $\sigma_\rho=1,2$:
$|\rho{\rangle_s}=|\sigma_\rho,\rho,l_\rho,M_\rho{\rangle_s}$.
The non-vanishing components of $\Gamma$ are
\be
&&\Gamma_{\mu\nu\rho}(G_\mu,+;G_\nu,+;l_\rho,1)=
\delta_{G_\mu G_\nu}\delta_{G_\mu l_\rho}
(2G_\mu+1)({\cal N}_\rho^{G_\mu})^2
\nonumber \\
&&\quad \times
\Big\{\sum_{n=1}^{2N(G_\mu)}\Big[\frac{G_\mu}{G_\mu+1}
V^{(+)}_{G_\mu}(n,\mu)V^{(+)}_{G_\mu}(n,\nu)-
V^{(+)}_{G_\mu}(n+2N(G_\mu),\mu)
V^{(+)}_{G_\mu}(n+2N(G_\mu),\nu)\Big]\Big\}
\nonumber \\
&&\quad \times
\int_0^D drr^2\int_0^D dr^\prime r^{\prime2}
\Big[a(r,\omega)a^*(r^\prime,\omega)
-b(r,\omega)b^*(r^\prime,\omega)\Big]
\label{gpp1} \\
&&\quad \times
\Big[{\overline w}_{\rho G_\mu}^{\hskip 0.3em-}j_{G_\mu}(k_{\rho G_\mu}r)
\tilde g_\nu^{(G_\mu,+;1)}(r)+
{\overline w}_{\rho G_\mu}^{\hskip 0.3em-}j_{G_\mu+1}(k_{\rho G_\mu}r)
\tilde f_\nu^{(G_\mu,+;1)}(r)\Big]
\nonumber \\
&&\quad\times
\Big[{\overline w}_{\rho G_\mu}^{\hskip 0.3em+}j_{G_\mu}(k_{\rho
G_\mu}r^\prime)
\tilde g_\mu^{(G_\mu,+;1)}(r^\prime)+
{\overline w}_{\rho G_\mu}^{\hskip 0.3em-}j_{G_\mu+1}(k_{\rho G_\mu}r^\prime)
\tilde f_\mu^{(G_\mu,+;1)}(r^\prime)\Big],
\nonumber
\vspace{0.5cm}
\\
&&\Gamma_{\mu\nu\rho}(G_\mu,+;G_\nu,+;l_\rho,2)=
-\delta_{G_\mu G_\nu}\delta_{G_\mu l_\rho}
(2G_\mu+1)({\cal N}_\rho^{G_\mu})^2
\nonumber \\
&&\quad \times
\Big\{\sum_{n=1}^{2N(G_\mu)}\Big[
V^{(+)}_{G_\mu}(n,\mu)V^{(+)}_{G_\mu}(n,\nu)-
\frac{G_\mu+1}{G_\mu}V^{(+)}_{G_\mu}(n+2N(G_\mu),\mu)
V^{(+)}_{G_\mu}(n+2N(G_\mu),\nu)\Big]\Big\}
\nonumber \\
&&\quad \times
\int_0^D drr^2\int_0^D dr^\prime r^{\prime2}
\Big[a(r,\omega)a^*(r^\prime,\omega)
-b(r,\omega)b^*(r^\prime,\omega)\Big]
\label{gpp2} \\
&&\quad \times
\Big[{\overline w}_{\rho G_\mu}^{\hskip 0.3em+}
j_{G_\mu}(k_{\rho G_\mu}r)
\tilde g_\nu^{(G_\mu,+;2)}(r)+
{\overline w}_{\rho G_\mu}^{\hskip 0.3em-}
j_{G_\mu-1}(k_{\rho G_\mu}r)
\tilde f_\nu^{(G_\mu,+;2)}(r)\Big]
\nonumber \\
&&\quad\times
\Big[{\overline w}_{\rho G_\mu}^{\hskip 0.3em+}
j_{G_\mu}(k_{\rho G_\mu}r^\prime)
\tilde g_\mu^{(G_\mu,+;2)}(r^\prime)+
{\overline w}_{\rho G_\mu}^{\hskip 0.3em-}
j_{G_\mu-1}(k_{\rho G_\mu}r^\prime)
\tilde f_\mu^{(G_\mu,+;2)}(r^\prime)\Big],
\nonumber
\vspace{0.5cm}
\\
&&\Gamma_{\mu\nu\rho}(G_\mu,+;G_\nu,-;l_\rho,1)=
\delta_{G_\mu G_\nu-1}\delta_{G_\mu l_\rho}
(2G_\mu+1)\frac{2G_\mu+3}{G_\mu+1}({\cal N}_\rho^{G_\mu})^2
\nonumber \\
&&\quad \times
\Big\{\sum_{n=1}^{2N(G_\mu)}V^{(+)}_{G_\mu}(n,\mu)
V^{(-)}_{G_\mu}(n+2N(G_\mu+2),\nu) \Big\}
\nonumber \\
&&\quad \times
\int_0^D drr^2\int_0^D dr^\prime r^{\prime2}
\Big[a(r,\omega)a^*(r^\prime,\omega)
-b(r,\omega)b^*(r^\prime,\omega)\Big]
\label{gpm1} \\
&&\quad \times
\Big[{\overline w}_{\rho G_\mu}^{\hskip 0.3em+}
j_{G_\mu}(k_{\rho G_\mu}r)
\tilde g_\nu^{(G_\mu+1,-;2)}(r)+
{\overline w}_{\rho G_\mu}^{\hskip 0.3em-}
j_{G_\mu+1}(k_{\rho G_\mu}r)
\tilde f_\nu^{(G_\mu+1,-;2)}(r)\Big]
\nonumber \\
&&\quad\times
\Big[{\overline w}_{\rho G_\mu}^{\hskip 0.3em+}
j_{G_\mu}(k_{\rho G_\mu}r^\prime)
\tilde g_\mu^{(G_\mu,+;1)}(r^\prime)+
{\overline w}_{\rho G_\mu}^{\hskip 0.3em-}
j_{G_\mu+1}(k_{\rho G_\mu}r^\prime)
\tilde f_\mu^{(G_\mu,+;1)}(r^\prime)\Big],
\nonumber \\
\vspace{0.5cm}
&&\Gamma_{\mu\nu\rho}(G_\mu,+;G_\nu,-;l_\rho,2)=
\delta_{G_\mu G_\nu+1}\delta_{G_\mu l_\rho}
(2G_\mu+1)\frac{2G_\mu-1}{G_\mu}({\cal N}_\rho^{G_\mu})^2
\nonumber \\
&&\quad \times
\Big\{\sum_{n=1}^{2N(G_\mu)}V^{(+)}_{G_\mu}(n+2N(G_\mu),\mu)
V^{(-)}_{G_\mu}(n,\nu) \Big\}
\nonumber \\
&&\quad \times
\int_0^D drr^2\int_0^D dr^\prime r^{\prime2}
\Big[a(r,\omega)a^*(r^\prime,\omega)
-b(r,\omega)b^*(r^\prime,\omega)\Big]
\label{gpm2} \\
&&\quad \times
\Big[{\overline w}_{\rho G_\mu}^{\hskip 0.3em+}
j_{G_\mu}(k_{\rho G_\mu}r)
\tilde g_\nu^{(G_\mu-1,-;1)}(r)+
{\overline w}_{\rho G_\mu}^{\hskip 0.3em-}
j_{G_\mu-1}(k_{\rho G_\mu}r)
\tilde f_\nu^{(G_\mu-1,-;1)}(r)\Big]
\nonumber \\
&&\quad\times
\Big[{\overline w}_{\rho G_\mu}^{\hskip 0.3em+}
j_{G_\mu}(k_{\rho G_\mu}r^\prime)
\tilde g_\mu^{(G_\mu,+;2)}(r^\prime)+
{\overline w}_{\rho G_\mu}^{\hskip 0.3em-}
j_{G_\mu-1}(k_{\rho G_\mu}r^\prime)
\tilde f_\mu^{(G_\mu,+;2)}(r^\prime)\Big],
\nonumber \\
\vspace{0.5cm}
&&\Gamma_{\mu\nu\rho}(G_\mu,-;G_\nu,+;l_\rho,1)=
\delta_{G_\mu G_\nu+1}\delta_{G_\mu l_\rho+1}
(2G_\mu+1)\frac{2G_\mu-1}{G_\mu}({\cal N}_\rho^{G_\mu-1})^2
\nonumber \\
&&\quad \times
\Big\{\sum_{n=1}^{2N(G_\mu-1)}V^{(-)}_{G_\mu}(n+2N(G_\mu+1),\mu)
V^{(+)}_{G_\mu-1}(n,\nu) \Big\}
\nonumber \\
&&\quad \times
\int_0^D drr^2\int_0^D dr^\prime r^{\prime2}
\Big[a(r,\omega)a^*(r^\prime,\omega)
-b(r,\omega)b^*(r^\prime,\omega)\Big]
\label{gmp1} \\
&&\quad \times
\Big[{\overline w}_{\rho G_\mu-1}^{\hskip 0.3em+}
j_{G_\mu-1}(k_{\rho G_\mu-1}r)
\tilde g_\nu^{(G_\mu-1,+;1)}(r)+
{\overline w}_{\rho G_\mu-1}^{\hskip 0.3em-}
j_{G_\mu}(k_{\rho G_\mu-1}r)
\tilde f_\nu^{(G_\mu-1,+;1)}(r)\Big]
\nonumber \\
&&\quad\times
\Big[{\overline w}_{\rho G_\mu-1}^{\hskip 0.3em+}
j_{G_\mu-1}(k_{\rho G_\mu-1}r^\prime)
\tilde g_\mu^{(G_\mu,-;2)}(r^\prime)+
{\overline w}_{\rho G_\mu-1}^{\hskip 0.3em-}
j_{G_\mu}(k_{\rho G_\mu-1}r^\prime)
\tilde f_\mu^{(G_\mu,-;2)}(r^\prime)\Big],
\nonumber
\vspace{0.5cm}
\\
&&\Gamma_{\mu\nu\rho}(G_\mu,-;G_\nu,+;l_\rho,2)=
\delta_{G_\mu G_\nu-1}\delta_{G_\mu l_\rho-1}
(2G_\mu+1)\frac{2G_\mu+3}{G_\mu+1}({\cal N}_\rho^{G_\mu+1})^2
\nonumber \\
&&\quad \times
\Big\{\sum_{n=1}^{2N(G_\mu+1)}V^{(-)}_{G_\mu}(n,\mu)
V^{(+)}_{G_\mu+1}(n+2N(G_\mu+1),\nu) \Big\}
\nonumber \\
&&\quad \times
\int_0^D drr^2\int_0^D dr^\prime r^{\prime2}
\Big[a(r,\omega)a^*(r^\prime,\omega)
-b(r,\omega)b^*(r^\prime,\omega)\Big]
\label{gmp2} \\
&&\quad \times
\Big[{\overline w}_{\rho G_\mu+1}^{\hskip 0.3em+}
j_{G_\mu+1}(k_{\rho G_\mu+1}r)
\tilde g_\nu^{(G_\mu+1,+;2)}(r)+
{\overline w}_{\rho G_\mu+1}^{\hskip 0.3em-}
j_{G_\mu}(k_{\rho G_\mu+1}r)
\tilde f_\nu^{(G_\mu+1,+;2)}(r)\Big]
\nonumber \\
&&\quad\times
\Big[{\overline w}_{\rho G_\mu+1}^{\hskip 0.3em+}
j_{G_\mu+1}(k_{\rho G_\mu+1}r^\prime)
\tilde g_\mu^{(G_\mu,-;1)}(r^\prime)+
{\overline w}_{\rho G_\mu+1}^{\hskip 0.3em-}
j_{G_\mu}(k_{\rho G_\mu+1}r^\prime)
\tilde f_\mu^{(G_\mu,-;1)}(r^\prime)\Big],
\nonumber \\
\vspace{0.5cm}
&&\Gamma_{\mu\nu\rho}(G_\mu,-;G_\nu,-;l_\rho,1)=
-\delta_{G_\mu G_\nu}\delta_{G_\mu l_\rho+1}
(2G_\mu+1)({\cal N}_\rho^{G_\mu-1})^2
\nonumber \\
&&\quad \times
\Big\{\sum_{n=1}^{2N(G_\mu+1)}V^{(-)}_{G_\mu}(n,\mu)
V^{(-)}_{G_\mu}(n,\nu)
\nonumber \\ &&\quad
-\frac{G_\mu+1}{G_\mu}
\sum_{n=1}^{2N(G_\mu-1)}V^{(-)}_{G_\mu}(n+2N(G_\mu+1),\mu)
V^{(-)}_{G_\mu}(n+2N(G_\mu+1),\nu) \Big\}
\nonumber \\
&&\quad \times
\int_0^D drr^2\int_0^D dr^\prime r^{\prime2}
\Big[a(r,\omega)a^*(r^\prime,\omega)
-b(r,\omega)b^*(r^\prime,\omega)\Big]
\label{gmm1} \\
&&\quad \times
\Big[{\overline w}_{\rho G_\mu-1}^{\hskip 0.3em+}
j_{G_\mu-1}(k_{\rho G_\mu-1}r)
\tilde g_\nu^{(G_\mu,-;2)}(r)+
{\overline w}_{\rho G_\mu-1}^{\hskip 0.3em-}
j_{G_\mu}(k_{\rho G_\mu-1}r)
\tilde f_\nu^{(G_\mu,-;2)}(r)\Big]
\nonumber \\
&&\quad\times
\Big[{\overline w}_{\rho G_\mu-1}^{\hskip 0.3em+}
j_{G_\mu-1}(k_{\rho G_\mu-1}r^\prime)
\tilde g_\mu^{(G_\mu,-;2)}(r^\prime)+
{\overline w}_{\rho G_\mu-1}^{\hskip 0.3em-}
j_{G_\mu}(k_{\rho G_\mu-1}r^\prime)
\tilde f_\mu^{(G_\mu,-;2)}(r^\prime)\Big],
\nonumber \\
\vspace{0.5cm}
&&\Gamma_{\mu\nu\rho}(G_\mu,-;G_\nu,-;l_\rho,2)=
\delta_{G_\mu G_\nu}\delta_{G_\mu l_\rho-1}
(2G_\mu+1)({\cal N}_\rho^{G_\mu+1})^2
\nonumber \\
&&\quad \times
\Big\{\sum_{n=1}^{2N(G_\mu+1)}
\frac{G_\mu}{G_\mu+1}V^{(-)}_{G_\mu}(n,\mu)
V^{(-)}_{G_\mu}(n,\nu)
\nonumber \\
&&\qquad
-\sum_{n=1}^{2N(G_\mu-1)}V^{(-)}_{G_\mu}(n+2N(G_\mu+1),\mu)
V^{(-)}_{G_\mu}(n+2N(G_\mu+1),\nu) \Big\}
\nonumber \\
&&\quad \times
\int_0^D drr^2\int_0^D dr^\prime r^{\prime2}
\Big[a(r,\omega)a^*(r^\prime,\omega)
-b(r,\omega)b^*(r^\prime,\omega)\Big]
\label{gmm2} \\
&&\quad \times
\Big[{\overline w}_{\rho G_\mu+1}^{\hskip 0.3em+}
j_{G_\mu+1}(k_{\rho G_\mu+1}r)
\tilde g_\nu^{(G_\mu,-;1)}(r)+
{\overline w}_{\rho G_\mu+1}^{\hskip 0.3em-}
j_{G_\mu}(k_{\rho G_\mu+1}r)
\tilde f_\nu^{(G_\mu,-;1)}(r)\Big]
\nonumber \\
&&\quad\times
\Big[{\overline w}_{\rho G_\mu+1}^{\hskip 0.3em+}
j_{G_\mu+1}(k_{\rho G_\mu+1}r^\prime)
\tilde g_\mu^{(G_\mu,-;1)}(r^\prime)+
{\overline w}_{\rho G_\mu+1}^{\hskip 0.3em-}
j_{G_\mu}(k_{\rho G_\mu+1}r^\prime)
\tilde f_\mu^{(G_\mu,-;1)}(r^\prime)\Big].
\nonumber
\ee

\smallskip
\leftline{\it C.4 The  Matrix elements involving $\tau_3
\times \tilde h_2(\omega,-\omega)$}

We finally list the matrix elements of the
type $\langle \mu |\tau_3|\nu \rangle\langle \nu
|\tilde h_{(2)}(\omega,-\omega) \mu \rangle$ which are needed to
complete the evaluation of the parameter $c$ in eqn. (\ref{cpar}).
Again we may perform the sum over the projection quantum
numbers. We therefore consider:
\be
\sum_{M_\mu, M_\nu}\langle \mu |\tau_3|\nu \rangle\langle \nu
|\tilde h_{(2)}(\omega,-\omega) \mu \rangle =
\frac{m+m_s}{24}\Delta_{\mu\nu}(G_\mu,\pi_\mu;G_\nu,\pi_\nu).
\label{defdelta}
\ee
We then obtain
\be
&&\Delta_{\mu\nu}(G_\mu,+;G_\nu,+)=(2G_\mu+1)\delta_{G_\mu G_\nu}
\int_0^D dr r^2
\Big(a(r,\omega)a^*(r,\omega)-b(r,\omega)b^*(r,\omega)\Big)
\nonumber \\
&&\quad \times
\Bigg\{\sum_{n=1}^{2N(G_\mu)}\Big[
V^{(+)}_{G_\mu}(n,\mu)V^{(+)}_{G_\mu}(n,\nu)
-\frac{G_\mu+1}{G_\mu}V^{(+)}_{G_\mu}(n+2N(G_\mu),\nu)
V^{(+)}_{G_\mu}(n+2N(G_\mu),\mu)\Big]\Bigg\}
\nonumber \\
&&\quad \times
\Bigg\{\frac{G_\mu}{G_\mu+1}\Big[
\tilde g_\nu^{(G_\mu,+;1)}(r)\tilde g_\mu^{(G_\mu,+;1)}(r)-
\tilde f_\nu^{(G_\mu,+;1)}(r)\tilde f_\mu^{(G_\mu,+;1)}(r)\Big]
\label{h2pp} \\ &&\qquad
-\Big[\tilde g_\nu^{(G_\mu,+;2)}(r)\tilde g_\mu^{(G_\mu,+;2)}(r)-
\tilde f_\nu^{(G_\mu,+;2)}(r)\tilde f_\mu^{(G_\mu,+;2)}(r)\Big]\Bigg\},
\nonumber \\
\vspace{0.5cm}
&&\Delta_{\mu\nu}(G_\mu,+;G_\nu,-)=(2G_\mu+1)
\int_0^D dr r^2
\Big(a(r,\omega)a^*(r,\omega)-b(r,\omega)b^*(r,\omega)\Big)
\nonumber \\
&&\quad \times
\Bigg\{\frac{2G_\mu+3}{G_\mu+1}\delta_{G_\mu+1 G_\nu}
\Big[\sum_{n=1}^{2N(G_\mu)}V^{(+)}_{G_\mu}(n,\mu)
V^{(-)}_{G_\mu+1}(n+2N(G_\mu+2),\nu)\Big]
\nonumber \\
&&\qquad \times
\Big[\tilde g_\nu^{(G_\mu+1,-;2)}(r)\tilde g_\mu^{(G_\mu,+;1)}(r)
-\tilde f_\nu^{(G_\mu+1,-;2)}(r)\tilde f_\mu^{(G_\mu,+;1)}(r)\Big]
\label{h2pm} \\
&&+\frac{2G_\mu-1}{G_\mu}\delta_{G_\mu-1 G_\nu}
\Big[\sum_{n=1}^{2N(G_\mu)}V^{(+)}_{G_\mu}(n+2N(G_\mu+2),\mu)
V^{(-)}_{G_\mu-1}(n,\nu)\Big]
\nonumber \\
&&\qquad \times
\Big[\tilde g_\nu^{(G_\mu-1,-;1)}(r)\tilde g_\mu^{(G_\mu,+;2)}(r)
-\tilde f_\nu^{(G_\mu-1,-;1)}(r)\tilde f_\mu^{(G_\mu,+;2)}(r)\Big]
\Bigg\},
\nonumber \\
\vspace{0.5cm}
&&\Delta_{\mu\nu}(G_\mu,-;G_\nu,+)=(2G_\mu+1)
\int_0^D dr r^2
\Big(a(r,\omega)a^*(r,\omega)-b(r,\omega)b^*(r,\omega)\Big)
\nonumber \\
&&\quad \times
\Bigg\{\frac{2G_\mu+3}{G_\mu+1}\delta_{G_\mu+1 G_\nu}
\Big[\sum_{n=1}^{2N(G_\mu+1)}V^{(-)}_{G_\mu}(n,\mu)
V^{(+)}_{G_\mu+1}(n+2N(G_\mu+1),\nu)\Big]
\nonumber \\
&&\qquad \times
\Big[\tilde g_\nu^{(G_\mu+1,+;2)}(r)\tilde g_\mu^{(G_\mu,-;1)}(r)
-\tilde f_\nu^{(G_\mu+1,+;2)}(r)\tilde f_\mu^{(G_\mu,-;1)}(r)\Big]
\label{h2mp} \\
&&+\frac{2G_\mu-1}{G_\mu}\delta_{G_\mu-1 G_\nu}
\Big[\sum_{n=1}^{2N(G_\mu-1)}V^{(-)}_{G_\mu}(n+2N(G_\mu+1),\mu)
V^{(+)}_{G_\mu-1}(n,\nu)\Big]
\nonumber \\
&&\qquad \times
\Big[\tilde g_\nu^{(G_\mu-1,+;1)}(r)\tilde g_\mu^{(G_\mu,-;2)}(r)
-\tilde f_\nu^{(G_\mu-1,+;1)}(r)\tilde f_\mu^{(G_\mu,-;2)}(r)\Big]
\Bigg\},
\nonumber \\
\vspace{0.5cm}
&&\Delta_{\mu\nu}(G_\mu,-;G_\nu,-)=(2G_\mu+1)\delta_{G_\mu G_\nu}
\int_0^D dr r^2
\Big(a(r,\omega)a^*(r,\omega)-b(r,\omega)b^*(r,\omega)\Big)
\nonumber \\
&&\quad \times
\Bigg\{\sum_{n=1}^{2N(G_\mu+1)}
\frac{G_\mu}{G_\mu+1} V^{(-)}_{G_\mu}(n,\mu)V^{(-)}_{G_\mu}(n,\nu)
\nonumber \\
&&\qquad
-\sum_{n=1}^{2N(G_\mu-1)}
V^{(-)}_{G_\mu}(n+2N(G_\mu+1),\nu)
V^{(-)}_{G_\mu}(n+2N(G_\mu+1),\mu)\Big]\Bigg\}
\nonumber \\
&&\quad \times
\Bigg\{\Big[\tilde g_\nu^{(G_\mu,-;1)}(r)\tilde g_\mu^{(G_\mu,-;1)}(r)-
\tilde f_\nu^{(G_\mu,-;1)}(r)\tilde f_\mu^{(G_\mu,-;1)}(r)\Big]
\label{h2mm} \\ &&\qquad
-\frac{G_\mu+1}{G_\mu}\Big[\tilde g_\nu^{(G_\mu,-;2)}(r)
\tilde g_\mu^{(G_\mu,-;1)}(r)-
\tilde f_\nu^{(G_\mu,-;2)}(r)\tilde f_\mu^{(G_\mu,-;2)}(r)\Big]\Bigg\}.
\nonumber
\ee
This completes the listing of all relevant matrix elements needed
to compute the parameters in the mass formula (\ref{mass}).

\vfill
\eject

\end{document}